\documentclass[aps,prx,reprint,superscriptaddress,floatfix,longbibliography]{revtex4-2}

\usepackage{amsmath}
\usepackage{amssymb}
\usepackage{graphicx}
\usepackage{bm}
\usepackage{xcolor}
\usepackage[colorlinks=true,linkcolor=blue,citecolor=blue,urlcolor=magenta]{hyperref}
\usepackage{upgreek}

\usepackage{cleveref}
\crefname{equation}{equation}{equations}
\Crefname{equation}{Equation}{Equations}
\crefname{figure}{Fig.}{Figs.}
\Crefname{figure}{Figure}{Figures}
\crefname{table}{Table}{Tables}
\Crefname{table}{Table}{Tables}
\renewcommand{\thetable}{\arabic{table}}  %
\crefname{section}{Sec.}{Secs.}
\Crefname{section}{Section}{Sections}
\crefname{subsection}{Sec.}{Secs.}
\Crefname{subsection}{Section}{Sections}
\crefname{subsubsection}{Sec.}{Secs.}
\Crefname{subsubsection}{Section}{Sections}

\newcommand{\subfigid}[1]{\textbf{#1}}

\newcommand{\wmod}{\omega_\mathrm{mod}}          %
\newcommand{\kex}{\kappa_\mathrm{ex}}            %
\newcommand{\kint}{\kappa_0}                     %
\newcommand{\Gopt}[1]{\Gamma_{\mathrm{opt},#1}}  %
\newcommand{\Gtot}[1]{\Gamma_{#1}}               %
\newcommand{\Gmulti}{\Gamma_\mathrm{multi}}      %
\newcommand{\Geff}[1]{\Gamma_{#1}^\mathrm{eff}}  %

\newcommand{\nth}[1]{n_{#1}^\mathrm{th}}
\newcommand{\neff}[1]{n_{#1}^\mathrm{eff}}
\newcommand{\nqba}[1]{n_{#1}^\mathrm{QBA}}       %
\newcommand{\nqbasm}{\bar{n}_\mathrm{QBA}}       %
\newcommand{\nss}{\bar{n}_\mathrm{steady}}

\newcommand{\chic}{\chi_\mathrm{c}}
\newcommand{\chieff}{\chi_\mathrm{eff}}

\newcommand{\uW}{\,\upmu\mathrm{W}}

\newcommand{\suppsec}[1]{Supplementary Note~\ref{#1}}
\newcommand{\suppfig}[2][]{Supplementary Fig.~\ref{#2}#1}
\newcommand{\maineqref}[1]{equation~(\ref{#1}) of the main text}
\newcommand{\mainfigref}[2][]{Fig.~\ref{#2}#1 of the main text}
\newcommand{\mainsecref}[1]{\hyperref[#1]{Methods} of the main text}

\begin{document}
\title{Reversal of dynamical backaction and cooling below the quantum backaction limit in Floquet-driven multimode optomechanical resonators}

\author{Menno H. Jansen}
\affiliation{Center for Nanophotonics, AMOLF, Amsterdam, The Netherlands}
\author{Roel Burgwal}
\altaffiliation[Present address: ]{QphoX B.V., Delft, The Netherlands}
\affiliation{Department of Applied Physics and Science Education, Eindhoven University of Technology, Eindhoven, The Netherlands}
\affiliation{Center for Nanophotonics, AMOLF, Amsterdam, The Netherlands}
\author{Ewold Verhagen}
\email{verhagen@amolf.nl}
\affiliation{Center for Nanophotonics, AMOLF, Amsterdam, The Netherlands}
\affiliation{Department of Applied Physics and Science Education, Eindhoven University of Technology, Eindhoven, The Netherlands}

\begin{abstract}
  Dynamical backaction describes how a driven cavity field responds with delay to mechanical motion, and can be used to cool that motion toward its quantum ground state or amplify it into self-oscillation. Control over this interaction underpins laser cooling, high-fidelity state preparation, quantum transduction, and phonon lasing. In conventional cavity optomechanics, both the sign and the magnitude of the optical damping are dictated by the laser detuning, and cooling is bounded by quantum backaction (QBA), the optomechanical analogue of the Doppler limit in cold atoms. We show, in theory and experiment, that dynamical backaction can be strongly enhanced in multimode optomechanical resonators driven by time-varying (Floquet) optical fields. A modulated drive couples a sideband-unresolved MHz mode to a sideband-resolved GHz mode through the shared cavity, replacing the QBA limit of the target mode by the occupancy of the auxiliary mode. Experimentally, we cool the MHz mode to $47 \pm 13$ phonons, about half its single-mode QBA limit and an order of magnitude below what single-tone backaction reaches at the same intracavity photon number. The character of the interaction can moreover be completely inverted at fixed laser detuning, turning cooling into amplification and vice versa, controlled solely by the modulation frequency. Floquet driving thus extends open-loop quantum control, which requires no measurement or feedback, to mechanical resonators outside the resolved-sideband regime.
\end{abstract}

\maketitle

\section{Introduction}
\label{sec:intro}

Cavity optomechanics makes use of radiation pressure to control mechanical motion via dynamical backaction\cite{kippenbergCavityOptomechanicsBackAction2008,aspelmeyer2014review}, a mechanism that governs mechanical dissipation and enables capabilities ranging from quantum state preparation\cite{teufelSidebandCoolingMicromechanical2011,chanLaserCoolingNanomechanical2011} to the realization of nanoscale phonon lasers\cite{kippenbergAnalysisRadiationPressureInduced2005,grudininPhononLaserAction2010} and study of chaos theory\cite{navarro-urriosNonlinearDynamicsChaos2017}. However, harnessing this backaction for quantum control is strictly constrained by fundamental limits. Just as discrete photon scattering imposes the Doppler limit on laser-cooling of atoms\cite{hanschCoolingGasesLaser1975,winelandLaserCoolingAtoms1979,diedrichLaserCoolingZeroPoint1989,monroeResolvedSidebandRamanCooling1995}, optical quantum fluctuations impose an analogous limit to laser cooling in optomechanics, known as the quantum backaction (QBA) limit\cite{marquardtQuantumTheoryCavityAssisted2007,wilson-raeTheoryGroundState2007}. This is brought about by quantum fluctuations of the drive field, which stimulate Stokes scattering. While ground-state cooling, i.e. thermal occupancies below unity, can be achieved in the resolved-sideband regime, these quantum fluctuations severely restrict cooling in the unresolved-sideband regime, typical of many macroscopic low-frequency resonators\cite{guoCoherentFeedbackOptomechanical2022}. Measurement-based feedback\cite{wilsonMeasurementbasedControlMechanical2015} sidesteps dynamical backaction and has cooled sideband-unresolved megahertz resonators below one phonon\cite{rossiMeasurementbasedQuantumControl2018,guoActivefeedbackQuantumControl2023}, but it requires a fast, precise measurement with high detection efficiency; coherent feedback, which replaces the measurement by an optical loop, has been proposed for the same regime\cite{guoCoherentFeedbackOptomechanical2022}. Overcoming the QBA limit of dynamical backaction itself would extend open-loop quantum control, which needs no measurement, to this broad class of resonators.

Expanding the system to multiple coupled modes offers a route beyond this limit, by giving access to rich collective dynamics with no single-mode analogue. Multimode optomechanics has recently driven breakthroughs across a broad spectrum of physics: it enables coherent, bidirectional quantum transduction between microwave and optical domains\cite{mirhosseiniSuperconductingQubitOptical2020, vanthielOpticalReadoutSuperconducting2025}, spontaneous mechanical synchronization\cite{zhangSynchronizationMicromechanicalOscillators2012, bagheriPhotonicCavitySynchronization2013}, and increased phonon laser stability\cite{mercadeFloquetPhononLasing2021}. It further enables nonreciprocal transport of both photons and phonons: multimode radiation pressure interactions can break time-reversal symmetry to realize magnetic-free optical isolation\cite{ruesinkNonreciprocityMagneticfreeIsolation2016} and on-chip optical circulation\cite{ruesinkOpticalCirculationMultimode2018}. The converse process, in which light mediates coupling between mechanical modes, breaks reciprocity for the mechanical degrees of freedom instead\cite{xuTopologicalEnergyTransfer2016,xuNonreciprocalControlCooling2019, delpinoNonHermitianChiralPhononics2022}. In levitated systems, tunable light-induced coupling between nanoparticles\cite{rieserTunableLightinducedDipoledipole2022} realizes nonreciprocal mechanical interactions and non-Hermitian collective dynamics\cite{reisenbauerNonHermitianDynamicsNonreciprocity2024}. In cavity-coupled solid-state resonators, where the coupling is mediated by a shared optical mode, this yields energy transfer whose outcome depends on the direction in which an exceptional point is encircled\cite{xuTopologicalEnergyTransfer2016} and chiral phonon transport\cite{delpinoNonHermitianChiralPhononics2022}. Broken reciprocity can even stimulate heat flow against thermal gradients\cite{xuNonreciprocalControlCooling2019, slimChiralThermalFluctuations2026}. In these demonstrations, and in experiments where two nearly degenerate modes share one cavity mode\cite{ockeloen-korppiSidebandCoolingNearly2019, dejongCoherentMechanicalNoise2022}, however, both mechanical modes lie in the same sideband regime and share essentially the same quantum backaction limit, so the coupling redistributes thermal energy between them rather than altering the limit either can reach.

It has been theoretically proposed that transitioning to a multimode architecture provides a workaround to the QBA limit\cite{ojanenGroundstateCoolingMechanical2014}. By parametrically coupling a target low-frequency mechanical mode to an auxiliary high-frequency mode, the effective dissipation can be modified. If higher optomechanical cooling rates can be achieved at equal intracavity photon fluctuations, this could allow reaching lower mechanical mode occupancies before quantum backaction is limiting. This again invites the analogy of atomic physics, where target ions that cannot be efficiently laser-cooled are instead sympathetically cooled by Coulomb-coupling them to a readily cooled auxiliary ion species\cite{larsonSympatheticCoolingTrapped1986}.

\begin{figure*}[tbp]
  \centering
  \includegraphics{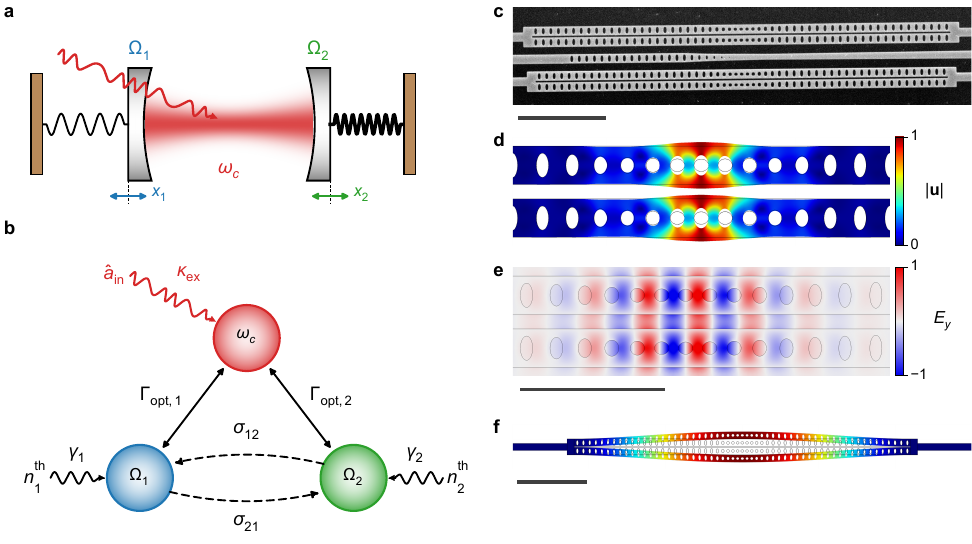}
  \caption{
    \textbf{Two mechanical modes in opposite sideband regimes share one optical mode.}
    \subfigid{a} Conceptual illustration: an optical cavity of resonance frequency $\omega_c$, driven by a laser, whose two mirrors are mechanical oscillators of frequencies $\Omega_1$ (blue, soft spring) and $\Omega_2$ (green, stiff spring) with displacements $x_1$ and $x_2$.
    \subfigid{b} The modes and their couplings: the drive $\hat{a}_\mathrm{in}$ enters the optical mode through $\kex$; the optical mode damps the two mechanical modes at $\Gopt{1}$ and $\Gopt{2}$, each of which is coupled to its thermal bath of occupancy $\nth{1}$, $\nth{2}$ at rate $\gamma_1$, $\gamma_2$. The dashed arrows are the optically induced cross-couplings $\sigma_{12}$ and $\sigma_{21}$.
    \subfigid{c} Scanning electron micrograph of the device.
    \subfigid{d} Simulated displacement profile $|\mathbf{u}|$ of the GHz mechanical mode $\Omega_2$.
    \subfigid{e} Simulated $y$ component of the electric field of the symmetric optical mode.    \subfigid{f} Simulated displacement profile of the MHz flexural mode $\Omega_1$, on the color scale of \subfigid{d}. All simulated fields are normalized to their maximum. Scale bars, $5~\upmu$m (\subfigid{c}, \subfigid{f}) and $2~\upmu$m (\subfigid{d}, \subfigid{e}, shared).
    \label{fig:intro}
  }
\end{figure*}

In this work, we realize coupling between two mechanical modes through a time-varying optical field, acting as the optomechanical equivalent of the Coulomb interaction by facilitating intermode coupling through a shared optical cavity, displayed schematically in \cref{fig:intro}a and b. We show how this radically alters the induced dynamics and lifts the standard quantum bounds on control. To this end, we use a platform of silicon photonic and phononic crystal resonators, which co-localizes high-frequency (GHz) and low-frequency (MHz) mechanical modes with a high-Q optical cavity, plotted in \cref{fig:intro}d-f, providing a versatile testbed for multimode dynamics. We demonstrate that this Floquet-driven intermode coupling both enhances dynamical backaction, cooling the low-frequency mode below its quantum backaction limit, and inverts its sign at fixed laser detuning, converting cooling into mechanical amplification and vice versa.

The coupled mechanical modes are first characterized through an effective dynamical matrix, whose eigenvalues yield the coupled mode frequencies and linewidths. This shows that modulation-induced coupling to a second mechanical mode produces much stronger changes to the mechanical damping rate of the low-frequency mechanical mode than in a single-mode system, at the same intracavity photon number. Solving the Lyapunov equation for this system yields the mode occupancies, and predicts both effects: cooling beyond the QBA limit, and on-demand switching of the sign of the backaction. We confirm both experimentally. Activating the intermode coupling between the target MHz mode and an auxiliary GHz mode cools the low-frequency resonator to about half of its radiation-pressure QBA limit, showing that Floquet-driven multimode dynamics surpass the single-mode cooling floor. At the exact same laser detuning, the system then transitions from cooling to mechanical amplification (gain), with the sign of the effect controlled solely by the frequency of the time-varying drive, producing a sharp and highly responsive transition between regimes.

\section{Results}
\label{sec:results}
\subsection{Review of single-mode dynamical backaction and quantum limit to cooling}
\label{subsec:res:singlemodetheory}

We first summarize single-mode optomechanical cooling theory, before extending it to our multimode system\cite{wilson-raeTheoryGroundState2007, marquardtQuantumTheoryCavityAssisted2007}. The canonical single-mode optomechanical system couples an optical cavity mode $\hat{a}$ (resonance frequency $\omega_c$, linewidth $\kappa$) to a mechanical resonator $\hat{b}$ (resonance frequency $\Omega_m$, linewidth $\gamma$) via radiation pressure at single-photon coupling rate $g_0$. The cavity is driven through an external port at rate $\kex$ and additionally suffers intrinsic loss at rate $\kint$, so that $\kappa = \kex + \kint$. Driving with a laser and linearizing around the resulting coherent amplitude $\alpha = \sqrt{\kex}\,\alpha_\mathrm{in}/(\kappa/2 - i\Delta)$, where $\Delta = \omega_L - \omega_c$ is the laser detuning, yields the equations of motion for the optical fluctuation operator $\hat{d} = \hat{a} - \alpha$ and the mechanical resonator\cite{aspelmeyer2014review}:

\begin{subequations}
  \begin{align}
    \dot{\hat{d}} &= -\left(\kappa/2 - i\Delta\right)\hat{d} - ig_0\alpha\left(\hat{b}+\hat{b}^\dagger\right) \nonumber\\
    &\quad + \sqrt{\kex}\,\hat{a}_\mathrm{in} + \sqrt{\kint}\,\hat{a}_{\mathrm{in},0}, \\
    \dot{\hat{b}} &= -\left(\gamma/2 + i\Omega_m\right)\hat{b} - ig_0\left(\alpha^*\hat{d} + \alpha\hat{d}^\dagger\right) + \sqrt{\gamma}\,\hat{\eta}.
  \end{align}
\end{subequations}

Transforming to the Fourier domain and adiabatically eliminating the optical field introduces an optomechanical modification to the mechanical susceptibility: $\Sigma(\omega) = |g_0\alpha|^2(\chic(\omega) - \chic^*(-\omega))$, where $\chic(\omega) = (\kappa/2 - i(\Delta+\omega))^{-1}$ is the optical susceptibility. In the weak-coupling regime ($g_0|\alpha| \ll \Omega_m$), applying the rotating-wave approximation yields an effective mechanical susceptibility $\chieff^{-1}(\omega) = \chi_m^{-1}(\omega) + \Sigma(\omega)$, where $\chi_m = (\gamma/2 - i(\omega - \Omega_m))^{-1}$. Evaluated at $\omega = \Omega_m$, the real and imaginary parts of $\Sigma$ give the optomechanical damping rate $\Gamma_\mathrm{opt} = 2\,\mathrm{Re}\{\Sigma(\Omega_m)\}$ and optical spring shift $\delta\Omega_m = \mathrm{Im}\{\Sigma(\Omega_m)\}$, respectively. This leads to the following expression for the optomechanical damping rate:

\begin{align}
  \Gamma_\mathrm{opt} = |g_0\alpha|^2 \biggl[&\frac{\kappa}{(\kappa/2)^2 + (\Delta+\Omega_m)^2} \nonumber\\
  &- \frac{\kappa}{(\kappa/2)^2+(\Delta-\Omega_m)^2}\biggr] = A_- - A_+.
  \label{eq:Gammaopt}
\end{align}

Here $A_-$ and $A_+$ are the anti-Stokes and Stokes scattering rates, respectively. Driving on the red sideband ($\Delta \approx -\Omega_m$) maximizes $A_- \gg A_+$, yielding a large positive $\Gamma_\mathrm{opt}$ that cools the resonator.

When $\Gamma_\mathrm{opt} \gg \gamma$, the resonator equilibrates with the optical field alone, and its occupancy is set by the balance between optical heating and cooling\cite{marquardtQuantumTheoryCavityAssisted2007}. The rate equation $\dot{\bar{n}} = -\Gamma_\mathrm{opt}\bar{n} + A_+$ then has a fixed point $\nss$, which is the lowest occupancy that radiation-pressure cooling can reach at a given detuning:

\begin{equation}
  \nss = \frac{A_+}{A_- - A_+}
  = \left(\frac{(\kappa/2)^2 + (\Delta-\Omega_m)^2}{(\kappa/2)^2 + (\Delta+\Omega_m)^2}-1\right)^{-1}.
  \label{eq:nsteady}
\end{equation}

By minimizing $\nss(\Delta)$, this reduces to the single-mode quantum backaction limit, which for the resolved-sideband regime ($\Omega_m \gg \kappa$), with optimal detuning $\Delta=-\Omega_m$, is given by\cite{wilson-raeTheoryGroundState2007, marquardtQuantumTheoryCavityAssisted2007}:

\begin{align}
  \nqbasm = \left(\frac{\kappa}{4\Omega_m}\right)^2,
  \label{eq:nqba-resolved}
\end{align}
while for the unresolved-sideband regime ($\Omega_m \ll \kappa$), with optimal detuning $\Delta=-\kappa/2$;

\begin{align}
  \nqbasm = \frac{\kappa}{4\Omega_m}.
  \label{eq:nqba-unresolved}
\end{align}

Ground-state cooling ($\bar{n}_\mathrm{min} \ll 1$) therefore requires the resolved-sideband condition $\kappa \ll 4\Omega_m$.

\subsection{Optically mediated coupling of a sideband-unresolved and a sideband-resolved mode}
\label{subsec:res:multimodetheory}
\begin{figure*}[tbp]
  \centering
  \includegraphics{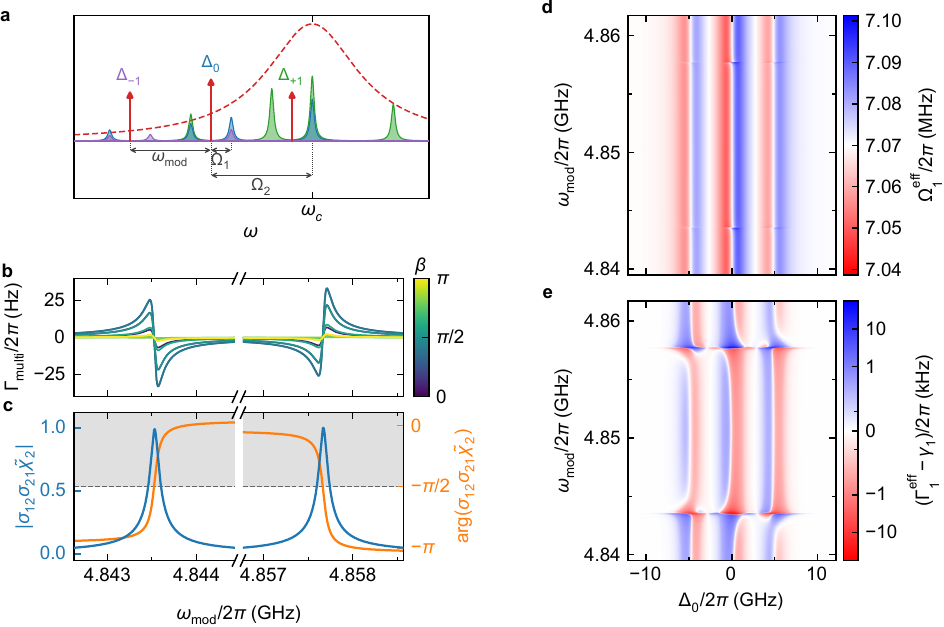}
  \caption{
    \textbf{A modulated drive overlaps the sidebands of the two modes inside the cavity, and the modulation frequency sets the sign of the resulting backaction.}
    \subfigid{a} Coupling scheme in the frequency domain, not to scale. Three drive tones (red arrows) at detunings $\Delta_{-1}$, $\Delta_0$, $\Delta_{+1}$ enter a cavity of susceptibility $|\chi(\omega)|^2$ (dashed red line), each generating sidebands of both mechanical modes (Lorentzians, colored by tone) offset by $\pm\Omega_1$ and $\pm\Omega_2$. At $\wmod = \Omega_2 \pm \Omega_1$, sidebands of different modes from neighboring tones become degenerate, mediating the intermode coupling $\sigma_{12}$, $\sigma_{21}$.
    \subfigid{b} Multimode contribution to the damping rate of mode 1, $\Gmulti$ (\cref{eq:multimode:Gmulti}), versus $\wmod$ for modulation depths $\beta$ from $0$ to $\pi$, at $\Delta_0 = -1.15\,\Omega_2$ and $0.5\uW$. Positive values are enhanced cooling, negative anti-damping.
    \subfigid{c} Magnitude (blue, left axis, normalized) and argument (orange, right axis) of $\sigma_{12}\sigma_{21}\tilde{\chi}_2(\Omega_1+\wmod)$ at $\beta = 0.4\pi$, otherwise as in \subfigid{b}. Sweeping $\wmod$ across the resonance rotates the phase by $\pi$; since $\Gmulti$ is set by the real part, the backaction reverses where the phase crosses $-\pi/2$ (dashed line; shading marks $\Gmulti < 0$).
    \subfigid{d}, \subfigid{e} Modified frequency (\subfigid{d}) and excess linewidth (\subfigid{e}) of mode 1 versus carrier detuning $\Delta_0$ and modulation frequency $\wmod$, at $\beta = 0.4\pi$ and $4.7\uW$. The color scale of \subfigid{e} is logarithmic beyond $\pm0.2$ kHz.
    \label{fig:res-theory:scheme-freq-space}
  }
\end{figure*}

Dynamics and limits can markedly differ if we consider a system of two mechanical resonators coupled to a common optical mode, extending the parameter regimes from previous theoretical treatments of multimode optomechanical systems\cite{ojanenGroundstateCoolingMechanical2014, xuTopologicalEnergyTransfer2016, xuNonreciprocalControlCooling2019, ockeloen-korppiSidebandCoolingNearly2019}. The two modes we consider occupy opposite sideband regimes: $\hat{b}_1$ is sideband-unresolved ($\Omega_1 \ll \kappa$) and $\hat{b}_2$ is sideband-resolved ($\Omega_2 > \kappa$). To bridge their large frequency separation, the cavity input field is phase-modulated with a modulation depth $\beta$, introducing three tones at detunings $\Delta_k = \Delta_0 + k\,\wmod$ ($k \in \{-1, 0, +1\}$) with amplitudes $\alpha_k = \sqrt{\kex}\,\alpha_{\mathrm{in},k}/(\kappa/2 - i\Delta_k)$, and $\alpha_{\mathrm{in},k} \propto J_k(\beta)$. Each tone independently generates mechanical sidebands in the optical fluctuation spectrum, indicated by the Lorentzians of various color in \cref{fig:res-theory:scheme-freq-space}a, with each color corresponding to mechanical sidebands of a particular tone. When two tones are separated by $\wmod \approx \Omega_2 \pm \Omega_1$, sidebands originating from different mechanical modes overlap spectrally inside the cavity, as illustrated in \cref{fig:res-theory:scheme-freq-space}a. Just below the frequency of the central $k=0$ laser tone we see its blue sideband from $\hat{b}_1^\dagger$ overlap with the green sideband of the $k=+1$ tone from $\hat{b}_2^\dagger$. This overlap, controlled by the cavity susceptibility, is the physical origin of the optomechanically induced coupling between the two modes.

To calculate the effect of this coupling, we linearize around the resulting large-amplitude optical state and adiabatically eliminate the optical field, similar to the single-mode case \cite{xuTopologicalEnergyTransfer2016,xuNonreciprocalControlCooling2019}. The full derivation is given in \suppsec{sec:SI:multimode}. We find that the modulated drive mediates an effective coupling between the two mechanical modes, with different nature depending on whether $\wmod$ is tuned to the sum or difference of the mode frequencies. The equations of motion in the rotating wave approximation take the form:

\begin{align}
  \dot{\hat{b}}_{1} &= -\left(\frac{\gamma_1}{2} + i\Omega_1
  + \sigma_{11}\right)\hat{b}_{1}
  - \sigma_{12}\,e^{i\wmod t}\,\hat{b}_{2} + \sqrt{\gamma_1}\,\hat{\eta}_1,
  \nonumber \\
  \dot{\hat{b}}_{2} &= -\left(\frac{\gamma_2}{2} + i\Omega_2
  + \sigma_{22}\right)\hat{b}_{2}
  - \sigma_{21}\,e^{-i\wmod t}\,\hat{b}_{1} + \sqrt{\gamma_2}\,\hat{\eta}_2,
  \label{eq:multimode:EOM}
\end{align}
where the optomechanical self-energies and cross-coupling rates are:

\begin{subequations}
  \begin{align}
    \sigma_{nn} &= g_n^2 \sum_{k=-1}^{+1} |\alpha_k|^2
    \left(\chi(\Omega_n + \Delta_k) - \chi(\Omega_n - \Delta_k)\right), \\
    \sigma_{12} &= g_1 g_2 \sum_{k=-1}^{0} \alpha_{k+1}^*\,\alpha_k
    \left(\chi(\Omega_1 + \Delta_{k+1}) - \chi(\Omega_1 - \Delta_k)\right), \\
    \sigma_{21} &= g_2 g_1 \sum_{k=-1}^{0} \alpha_k^*\,\alpha_{k+1}
    \left(\chi(\Omega_2 + \Delta_k) - \chi(\Omega_2 - \Delta_{k+1})\right),
  \end{align}
  \label{eq:multimode:sigmas}
\end{subequations}
with $\chi(\omega) = (\kappa/2 - i\omega)^{-1}$, so that $\chic(\omega) = \chi(\Delta + \omega)$ and $\chi(-\omega) = \chi^*(\omega)$. The diagonal terms $\sigma_{nn}$ generalize the single-mode self-energy $\Sigma(\Omega_m)$ to three driving tones, and reduce to it for $\beta \to 0$. The off-diagonal terms $\sigma_{12}$ and $\sigma_{21}$ represent the optomechanically mediated coupling between the two mechanical modes.

To see how this coupling acts on mode 1, we solve \cref{eq:multimode:EOM} in the frequency domain and eliminate $\hat{b}_2$. Ignoring laser noise for now, the spectral amplitude of mode 1 reads, for $\wmod \approx \Omega_2 - \Omega_1$,

\begin{equation}
  \begin{split}
  &\hat{b}_{1}[\omega] = \frac{\tilde{\chi}_{1}(\omega)}{1 - \tilde{\chi}_{1}(\omega)\,\tilde{\chi}_{2}(\omega+\wmod)\,\sigma_{12}\sigma_{21}}\\
  &\quad\times\bigl[\sqrt{\gamma_{1}}\,\hat{\eta}_{1}[\omega] - \sigma_{12}\,\tilde{\chi}_{2}(\omega+\wmod)\sqrt{\gamma_{2}}\,\hat{\eta}_{2}[\omega+\wmod]\bigr],
  \end{split}
  \label{eq:multimode:b1spectrum}
\end{equation}
where $\tilde{\chi}_n(\omega) = (\gamma_n/2 - i(\omega - \Omega_n) + \sigma_{nn})^{-1}$ is the optomechanically modified susceptibility of mode $n$. The denominator captures fluctuations that couple from mode 1 into mode 2 and act back on mode 1: the susceptibility of mode 1 to its own bath is renormalized by the interaction, $\tilde{\chi}_{1,\mathrm{eff}}^{-1} = \tilde{\chi}_{1}^{-1} - \sigma_{12} \sigma_{21} \tilde{\chi}_{2}$, so that the total damping of mode 1 becomes $\Geff{1} = \Gtot{1} + \Gmulti$. Here $\Gtot{n} = \gamma_n + \Gopt{n}$ is the damping that mode $n$ has without the multimode coupling, with $\Gopt{n} = 2\,\mathrm{Re}\,\sigma_{nn}$ its optomechanical damping by the three tones, and the multimode contribution is

\begin{equation}
  \Gmulti = -2\,\mathrm{Re}\!\left[\sigma_{12}\sigma_{21}\,\tilde{\chi}_2(\Omega_1 + \wmod)\right].
  \label{eq:multimode:Gmulti}
\end{equation}

$\Gmulti$ is shown in \cref{fig:res-theory:scheme-freq-space}b as a function of $\wmod$ at fixed carrier detuning $\Delta_0 = -1.15\,\Omega_2$, chosen so that the two lobes of the reversal are comparable in magnitude, and an input power of $0.5\uW$ at the device, for several modulation depths $\beta$. Two resonances appear, at $\wmod \approx \Omega_2 - \Omega_1$ and $\wmod \approx \Omega_2 + \Omega_1$, precisely where the sideband overlap of \cref{fig:res-theory:scheme-freq-space}a occurs. At each of them $\Gmulti$ changes sign, giving enhanced cooling on one side and anti-damping on the other. The damping of mode 1 is therefore altered, and its sign reversed, by tuning $\wmod$ alone: the carrier detuning, the total optical power, and the optomechanical coupling rates are all held fixed across the sweep, and only the spectral degeneracy condition is changed.

The origin of the reversal is apparent in \cref{fig:res-theory:scheme-freq-space}c, which shows the magnitude and argument of the product $\sigma_{12}\sigma_{21}\tilde{\chi}_2(\Omega_1+\wmod)$ entering \cref{eq:multimode:Gmulti}. The magnitude is resonant, turning the coupling on over a bandwidth set by the total linewidth of the sideband-resolved mode, $\Gtot{2} = \gamma_2 + \Gopt{2} \approx 2\pi\times100$ kHz. Across that same span the argument rotates by $\pi$, and since $\Gmulti$ is set by the real part, the backaction (i.e. the added damping rate) reverses when the phase crosses $-\pi/2$. So here we recognize that the high-frequency mechanical mode's response is responsible for the retardation that is essential to dynamical backaction. The modulation frequency thus controls the sign of the multimode backaction in the same way that the optical detuning controls the sign of single-mode dynamical backaction, but over a bandwidth smaller than the optical linewidth $\kappa$ by more than four orders of magnitude, rather than one set by $\kappa$ itself.

Importantly, \cref{fig:res-theory:scheme-freq-space}b shows that $|\Gmulti|$ can be twice the single-mode value $|\Gopt{1}| = 2\pi\times15$ Hz at the same intracavity photon number, for modulation depths near $\beta = 0.3\pi$, close to the optimum depth discussed below. The interaction with the second mode introduces the possibility to boost the retardation of the backaction, greatly enhancing dissipation modifications.

The full control space is mapped in \cref{fig:res-theory:scheme-freq-space}d and e, which show the modified frequency and linewidth of mode 1 over a simultaneous sweep of $\Delta_0$ and $\wmod$, at an increased input power of $4.7\uW$ at the device, that of the measurement in \cref{fig:res-exp:freq-lw-n1eff-mainsweep}. The single-mode backaction contained in $\sigma_{11}$ depends only on $\Delta_0$, producing broad bands that reverse sign as the tone sweeps across the optical resonance, repeated once for each of the three drive tones. The multimode coupling depends also on $\wmod$, producing the sharp asymmetric features at $\wmod \approx \Omega_2 \pm \Omega_1$ that cut across those bands. That the multimode response runs orthogonal to the detuning axis is exactly the statement that it acts as a control parameter independent of the laser detuning: at both positive and negative $\Delta_0 \approx \pm \Omega_2$, the modulation frequency determines whether the coupling adds damping or removes it, regardless of the sign of the single-mode backaction.

\subsection{Occupancies and the multimode cooling floor}
The $\hat{\eta}_2$ term of \cref{eq:multimode:b1spectrum} shows fluctuations of mode 2 acting directly on mode 1. This sets the floor of the scheme: as the multimode coupling is increased, mode 1 is driven toward the steady-state occupancy of mode 2 rather than toward zero. We quantify this floor below, once the steady-state occupancies have been obtained.

To derive an expression for the mechanical occupancies, we collect the mechanical amplitudes in $\bm{b} = (\hat{b}_1, \hat{b}_2)^T$ and move to a frame co-rotating at $\pm\wmod/2$, defined by $\bm{b}^{r}(t) = U(t)\bm{b}(t)$ with $U(t) = \mathrm{diag}(e^{-i\wmod t/2},\, e^{i\wmod t/2})$. In this frame both modes appear at $(\Omega_1 + \Omega_2)/2$. Once the rotating-wave approximation has removed the terms that oscillate at $\wmod$ and $2\wmod$, the coefficients of the matrix equations of motion $\dot{\bm{b}}^{r}(t) = D^{\mathrm{rot}} \bm{b}^{r}(t) + \bm{f}^{r}(t)$ are constant in this frame, with dynamical matrix\cite{jiangNonreciprocalDynamicsCryogenic2020}

\begin{align}
  \label{eq:multimode:Drot}
  \begin{split}
  D^{\mathrm{rot}} ={}& -i\begin{pmatrix}
    \Omega_1 + \wmod/2 & 0 \\
    0 & \Omega_2-\wmod/2
  \end{pmatrix}\\
  &- \begin{pmatrix}
    \gamma_1/2 + \sigma_{11} & \sigma_{12} \\
    \sigma_{21} & \gamma_2/2 + \sigma_{22}
  \end{pmatrix},
  \end{split}
\end{align}
where the signs of $\wmod/2$ apply for modulation near the difference frequency $\wmod \approx \Omega_2 - \Omega_1$. The sum-frequency case $\wmod \approx \Omega_2 + \Omega_1$ couples $\hat{b}_1$ to $\hat{b}_2^\dagger$, which appears at negative frequency in the rotating frame, with an inverted energy ladder like that of the effective negative-mass mode used to invert dynamical backaction in a superconducting photon-pressure circuit\cite{rodriguesPhotonPressureEffective2024}. It yields similar effects, with the cooling and anti-damping lobes mirrored, and is treated in \suppsec{subsec:SI:sumfreq}. The eigenvalues of $D^\mathrm{rot}$ take the form $\lambda_\pm = -\Geff{\pm}/2 - i\Omega^{\mathrm{eff}}_{\pm}$, and give the linewidths $\Geff{\pm}$ and frequencies $\Omega^{\mathrm{eff}}_{\pm}$ of the hybridized modes in the rotating frame. The corresponding laboratory-frame frequencies follow by undoing the shift of $\mp\wmod/2$.

The steady-state phonon occupancy of each mode is obtained by solving the Lyapunov equation for the covariance matrix, $V_{nm} = \langle \hat{b}_m^\dagger \hat{b}_n\rangle$\cite{serafiniQuantumContinuousVariables2023}:

\begin{equation}
  \label{eq:multimode:lyapunov}
  D^{\mathrm{rot}} V + V \left(D^{\mathrm{rot}}\right)^\dagger + Q = 0,
\end{equation}
where the noise diffusion matrix $Q$ accounts for both heating from the mechanical bath and optical quantum fluctuations, including cross-correlated quantum backaction arising from the fact that both modes are driven by the same intracavity field:

\begin{equation}
  \label{eq:multimode:Q}
  Q =
  \begin{pmatrix}
    \gamma_1 \nth{1} + A_{+,1} & A_{\times,12} \\
    \left(A_{\times,12}\right)^* &\gamma_2 \nth{2} + A_{+,2}
  \end{pmatrix}.
\end{equation}

\begin{figure*}[tbp]
  \centering
  \includegraphics{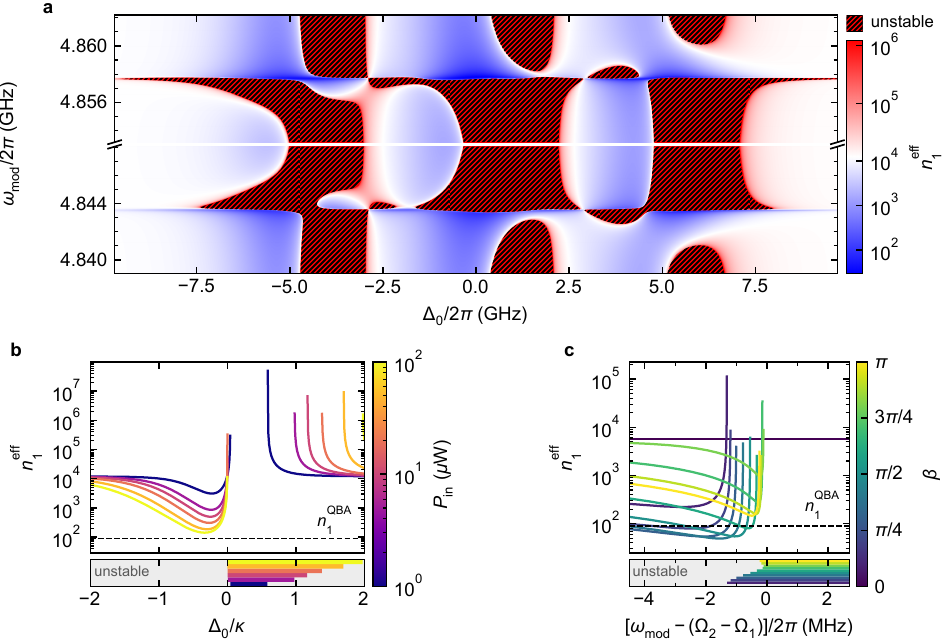}
  \caption{
    \textbf{The model predicts cooling below the single-mode quantum backaction limit, which a single tone never reaches.}
    \subfigid{a} Calculated effective occupancy of mode 1, $\neff{1}$, as a function of carrier detuning and modulation frequency, for an input power of $10\uW$ and a modulation depth of $0.4\pi$. The equilibrium phonon number $\nth{1}=1.2\times10^4$ is indicated by white, with red indicating heating and blue indicating cooling. Regions in which either mode is dynamically unstable admit no steady state and are hatched.
    \subfigid{b} The effective occupancy of mode 1 with a single unmodulated input laser tone, as a function of detuning, for various input powers. The unstable blue-detuned region is indicated by bars in the bottom panel. The dashed line indicates the single-mode QBA limit $\nqba{1}=88.9$.
    \subfigid{c} Floquet-cooled occupancy of mode 1, as a function of modulation detuning from the difference frequency, for various modulation depths at $\Delta_0=-\Omega_2$ and an input power of $100\uW$. The dashed line indicates the single-mode QBA limit $\nqba{1}$, showing that multimode cooling can surpass it.
    \label{fig:res-theory:n1eff-single-vs-multi}
  }
\end{figure*}

Here $\nth{n} = k_\mathrm{B}T/\hbar\Omega_n$ is the thermal occupancy. The optical heating rates $A_{+,n}$ and cross-backaction term $A_{\times,12}$ originate from the optical vacuum noise force after adiabatic elimination of the intracavity field. Both the external coupling port ($\hat{a}_\mathrm{in}$, rate $\kex$) and the intrinsic loss port ($\hat{a}_{\mathrm{in},0}$, rate $\kint$) inject independent vacuum fluctuations into the cavity. Because these ports are statistically independent, their contributions to the force noise power add incoherently, yielding a combined prefactor $\kex + \kint = \kappa$ in all optical heating terms. The per-mode heating rate is therefore

\begin{equation}
  \label{eq:multimode:Aplus}
  A_{+,n} = g_n^2\sum_{k=-1}^{+1} |\alpha_k|^2
  \frac{\kappa}{(\kappa/2)^2 + (\Delta_k - \Omega_n)^2},
\end{equation}
which is the total optical heating rate of mode $n$ from all three tones, consistent with the single-mode result $A_+ = |g_0\alpha|^2\kappa\,((\kappa/2)^2 + (\Delta-\Omega_m)^2)^{-1}$. The off-diagonal element is the optical cross-diffusion, which for modulation near the difference frequency reads

\begin{equation}
  \label{eq:multimode:Across}
  A_{\times,12} = g_1 g_2\,\kappa \sum_{k=-1}^{0} \alpha_k\,\alpha_{k+1}^*\,
  \chi(\Omega_1 - \Delta_k)\,\chi^*(\Omega_2 - \Delta_{k+1}).
\end{equation}
This term represents optical vacuum fluctuations interacting with both modes via the same intracavity photon, which requires that photon to be a mechanical sideband of both. Because $\Omega_1 \neq \Omega_2$, no single tone provides such an overlap; it arises only when the sideband of tone $k$ on one mode coincides with the sideband of tone $k+1$ on the other, the same degeneracy that generates the coupling $\sigma_{12}$ in \cref{fig:res-theory:scheme-freq-space}a. The cross-diffusion is therefore built from the adjacent-tone products $\alpha_k\alpha_{k+1}^*$ rather than from the tone powers $|\alpha_k|^2$ that enter $A_{+,n}$. The size of $A_{\times,12}$ is set by the cavity filtering of the shared photon. For red-detuned tones near the difference frequency that photon is the Stokes sideband of both modes, at $-(\Omega_1+\Omega_2)$ from cavity resonance, and since $\Omega_1 \ll \Omega_2$ the cross-diffusion is suppressed relative to the diagonal terms by
\begin{equation}
  \epsilon \equiv \frac{|\chi(\Omega_2)|^2}{|\chi(0)|^2} = \frac{(\kappa/2)^2}{(\kappa/2)^2 + \Omega_2^2} \approx 0.063.
  \label{eq:multimode:epscross}
\end{equation}
The same factor returns in the cooling limits, wherever a tone scatters a photon about $\Omega_2$ away from the cavity resonance. Near the sum frequency the shared photon lies at cavity resonance, and the cross-diffusion is of the same order as the coupling itself, $|A_{\times,12}| \sim |\sigma_{12}|$, with the consequences for the cooling floor discussed below. The diagonal elements $V_{11}$ and $V_{22}$ of the solution directly yield the effective phonon occupancies $\neff{1}$ and $\neff{2}$. In the limit of a single tone and no cross-coupling, \cref{eq:multimode:lyapunov} reduces to the single-mode result of \cref{eq:nqba-unresolved}, $\nqbasm = \kappa/4\Omega_1$, at the optimal detuning $\Delta_0 = -\kappa/2$. This can be seen in \cref{fig:res-theory:n1eff-single-vs-multi}b, which shows the result of solving \cref{eq:multimode:lyapunov} for $\neff{1}$ in the case of a single unmodulated input laser tone. For increasing input power the minimum occupancy approaches the QBA limit (the dashed line), but never surpasses it.

We solve \cref{eq:multimode:lyapunov} for $\neff{1}$ over a grid of modulation frequencies and carrier detunings at an input power of $10\uW$ and a modulation depth of $0.4\pi$, starting from a bath temperature of 4.1 K, corresponding to a thermal occupancy $\nth{1}=1.2 \times 10^4$. The resulting effective occupancies are shown in \cref{fig:res-theory:n1eff-single-vs-multi}a. Two features stand out. First, there are three broad regions, at modulation frequencies away from $\Omega_2 \pm \Omega_1$, in which one of the two modes becomes dynamically unstable as the laser tones scan across the cavity; there the Lyapunov equation admits no steady state and the corresponding pixels are masked. Second, the effect of the coupling on mode 1 changes sharply on crossing the difference and sum frequencies, producing both enhanced cooling and strong mechanical gain. This is in line with the expectations from \cref{fig:res-theory:scheme-freq-space}b-e.

At the input power used here, the single-tone calculation of \cref{fig:res-theory:n1eff-single-vs-multi}b gives a minimum of $\neff{1} = 500$ at the optimal detuning, more than five times larger than $\nqba{1}$. Activating the modulation, the lowest occupancy in \cref{fig:res-theory:n1eff-single-vs-multi}a is $\neff{1} = 29$, more than two orders of magnitude below the thermal occupancy and well below $\nqba{1} = 88.9$. The two calculations share the input power, the optomechanical coupling rates and the bath temperature, and differ only in whether the modulation sidebands are present. The difference of more than an order of magnitude between them is therefore attributable entirely to the additional dissipation channel opened by the auxiliary mode.

Finally, \cref{fig:res-theory:n1eff-single-vs-multi}c depicts $n_1^\mathrm{eff}$ for a fixed carrier detuning $\Delta_0=-\Omega_2$ and the maximum input power used in \cref{fig:res-theory:n1eff-single-vs-multi}b, 100 $\upmu$W, for various modulation depths, as a function of the modulation detuning from the difference frequency. On increasing the modulation depth the minimum occupancy drops below the single-mode QBA limit before rising again, showing that the multimode interaction improves the cooling performance beyond what is possible in a single-mode system. The existence of an optimum modulation depth follows from the structure of $\sigma_{12}$. Both terms in the sum of \cref{eq:multimode:sigmas} carry a product proportional to $J_0(\beta)J_1(\beta)$. The coupling strength therefore scales as $J_0(\beta)J_1(\beta)$, which is maximal at $\beta \approx 0.34\pi$.

The floor that \cref{fig:res-theory:n1eff-single-vs-multi}c approaches follows from \cref{eq:multimode:lyapunov} in the limit $\Gmulti \gg \Gtot{1}$, which we derive in \suppsec{subsec:SI:coolinglimit}:

\begin{equation}
  \neff{1} \;\longrightarrow\;
  \frac{A_{+,1}}{\Gmulti} +
  \frac{|\sigma_{12}|}{|\sigma_{21}|}\,\neff{2},
  \label{eq:multimode:floor}
\end{equation}
with $\Gmulti$ evaluated at the maximum of its cooling lobe. The QBA limit of the target mode is thus replaced by the occupancy of the auxiliary mode, weighted by the ratio $|\sigma_{12}|/|\sigma_{21}|$ that measures the reciprocity of the coupling, plus a residual from the Stokes heating of mode 1 by its own tone. Near the sum frequency the cross-diffusion $A_{\times,12}$ interferes destructively with the mode-2 noise transmitted through $\sigma_{12}$; this removes the residual together with the vacuum contribution of the auxiliary mode and leaves $\neff{1} \to \neff{2} + \gamma_2/\Gopt{2}$, up to small off-resonant scattering terms, as we derive in \suppsec{subsec:SI:coolinglimit-sum}. This is the interference proposed for ground-state cooling in the unresolved-sideband regime \cite{ojanenGroundstateCoolingMechanical2014}, which the Floquet scheme contains as one of its two branches.

The auxiliary mode is itself damped by the carrier tone, so $\neff{2}$ is set by the balance between its thermal load, its own optical damping, and the heat it receives from mode 1 through the coupling. At a bath temperature of $4.1$ K and the input power of the measurement reported below, $14.0\uW$, mode 2 is thermally limited rather than backaction-limited. At $\Delta_0 = -\Omega_2$, which places the carrier on the red sideband of the auxiliary mode, and at $\wmod = \Omega_1 + \Omega_2$, $\neff{2} \approx 18$; we use this as the reference point of the model. This places the floor for mode 1 at about $20$ phonons, a factor of four below the single-mode limit $\nqba{1} = \kappa/4\Omega_1 = 88.9$ of the MHz mode. Including the thermal load of mode 1 itself, the full solution of the Lyapunov equation gives $\neff{1} = 34$ there, as we show in \suppsec{subsec:SI:coolinglimit-sum}. With a colder bath or a stronger carrier the sum-frequency floor follows $\neff{2}$ down to a few phonons.

\subsection{Backaction reversal and cooling below the quantum backaction limit}
\label{subsec:res:experiment}
We realize this multimode interaction scheme in a photonic crystal nanobeam zipper cavity\cite{chanOpticalMechanicalDesign2009}, containing both a MHz and a GHz mechanical mode, coupled to the same optical mode. This mode is driven by a strong modulated laser, as described above. The effect of this drive is recorded by an additional weak probe laser, with its frequency locked to a secondary optical mode of the device. This laser measures the mechanical fluctuation spectra, used to determine the changing frequencies, linewidths, and occupancies as the drive laser parameters are varied. Details of the experimental setup are given in Methods. The input powers quoted in this work are the monitored power multiplied by the measured efficiencies of the setup, and are therefore powers at the device. The device is displayed in \cref{fig:intro}c-f, and its parameters are listed in \cref{methtab:dev-params}.

\begin{figure*}[tbp]
  \centering
  \includegraphics{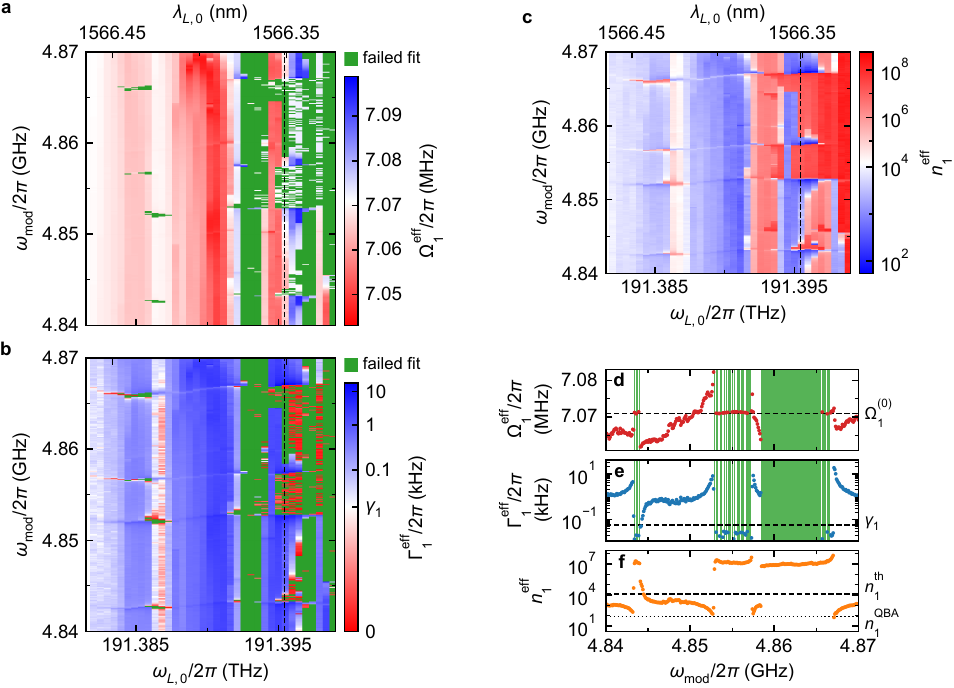}
  \caption{
    \textbf{Measured backaction reverses on crossing the resonance condition, at fixed laser detuning.}
    \subfigid{a}--\subfigid{c} The frequency $\Omega^{\mathrm{eff}}_1/2\pi$, linewidth $\Geff{1}/2\pi$, and occupancy $\neff{1}$ of mode 1 as a function of the frequency of the central laser tone (its wavelength $\lambda_{L,0}$ on the top axis) and the modulation frequency, for an input power of $4.7\uW$ at the device and a modulation depth of $0.45\pi$. White marks the unperturbed value in all three panels. Green pixels in \subfigid{a} and \subfigid{b} mark spectra for which the Lorentzian fit did not converge because the mode is self-oscillating.
    \subfigid{d}--\subfigid{f} Slices of \subfigid{a}--\subfigid{c} along the constant laser frequency $\omega_{L,0}/2\pi=191.3954$ THz indicated by the dashed lines, with green bands where the fit did not converge. Dashed lines mark the bare $\Omega_1$, $\gamma_1$, and $\nth{1}=1.2\times10^4$, the dotted line $\nqba{1}=88.9$.
    \label{fig:res-exp:freq-lw-n1eff-mainsweep}
  }
\end{figure*}

We sweep the frequency of the carrier laser tone $\omega_{L,0}/2\pi$ and the modulation frequency $\wmod/2\pi$, and at each combination record the spectrum of mode 1. We fit a Lorentzian to the fluctuation spectra to extract the resonance frequency $\Omega^{\mathrm{eff}}_1/2\pi$ and total linewidth $\Geff{1}/2\pi$, which are shown in \cref{fig:res-exp:freq-lw-n1eff-mainsweep}a and b respectively, for an input power of $4.7\uW$ and a modulation depth of $0.45\pi$. When the mechanical resonator is self-oscillating, its spectrum is no longer Lorentzian, such that the fit does not converge. These points are indicated by the green pixels in \cref{fig:res-exp:freq-lw-n1eff-mainsweep}a and b. By summing the power in the signal after subtracting the noise floor, and calibrating it against a reference spectrum taken with the driving laser far off-resonance, where the mode is in equilibrium with the $4.1$ K cold finger and the weak probe heats it by only about $0.01$ K, we obtain the occupancy of the mode at each drive detuning and modulation frequency, which is depicted in \cref{fig:res-exp:freq-lw-n1eff-mainsweep}c. The calibration and its checks are described in Methods.

In \cref{fig:res-exp:freq-lw-n1eff-mainsweep} there are two sets of difference and sum frequencies, because two high-frequency modes take the role of mode 2 of the two-mode theory. On top of the background of dynamical backaction from the three laser tones, we observe the predicted switches in the sign of the optical damping rate from \cref{fig:res-theory:scheme-freq-space}b and e at 4 frequencies, $\sim 4.850\pm0.007$ and $\sim 4.859\pm0.007$ GHz in \cref{fig:res-exp:freq-lw-n1eff-mainsweep}b, at both positive and negative carrier detuning. We note that the predicted broad bands of self-oscillation are only present on the blue-detuned side. This could be due to photothermal backaction\cite{primoAccurateModelingCharacterization2021,jansenControllingPhotothermalForces2026}, which is not included in our current model, limiting the region where a negative effective linewidth is attained.

As the determination of the occupancy does not rely on a fitting procedure, it remains accurate even when the mode is self-oscillating, which is why \cref{fig:res-exp:freq-lw-n1eff-mainsweep}c is cleaner than panels a and b. It shows the enhancement of the cooling when the modulation brings the sidebands of modes 1 and 2 close to resonance, and clearly resolves the abrupt switch from enhanced cooling to gain at fixed carrier detuning. The resonance condition for $\wmod$ drifts slightly with carrier detuning, following the optical spring shift of mode 2.

Panels d-f of \cref{fig:res-exp:freq-lw-n1eff-mainsweep} show a cut through the sweep at the fixed carrier frequency $\omega_{L,0}/2\pi = 191.3954$ THz, indicated by the dashed lines in panels a-c. The cut crosses both an enhanced-cooling region and a self-oscillation band, so a single trace spans the full range of the effect. The frequency shift in panel d is of order $10^{-3}\,\Omega_1$ and changes sign across the resonance condition, as expected for the imaginary part of the multimode interaction-induced self-energy. The linewidth in panel e varies over more than two decades, reaching values above $10$ kHz where the sidebands are brought close to resonance by the modulation, and falling below the intrinsic $\gamma_1$ on the other side of the resonance condition where the multimode contribution to the damping is negative. The green bands mark the modulation frequencies at which the mode self-oscillates and the Lorentzian fit does not converge. Because the occupancy is obtained from the integrated signal power rather than from a fit, panel f is continuous across those bands. It falls to roughly two orders of magnitude below the thermal occupancy $\nth{1}$ at the cooling optimum, and rises by three further decades inside the self-oscillation region.

\begin{figure*}[tbp]
  \centering
  \includegraphics{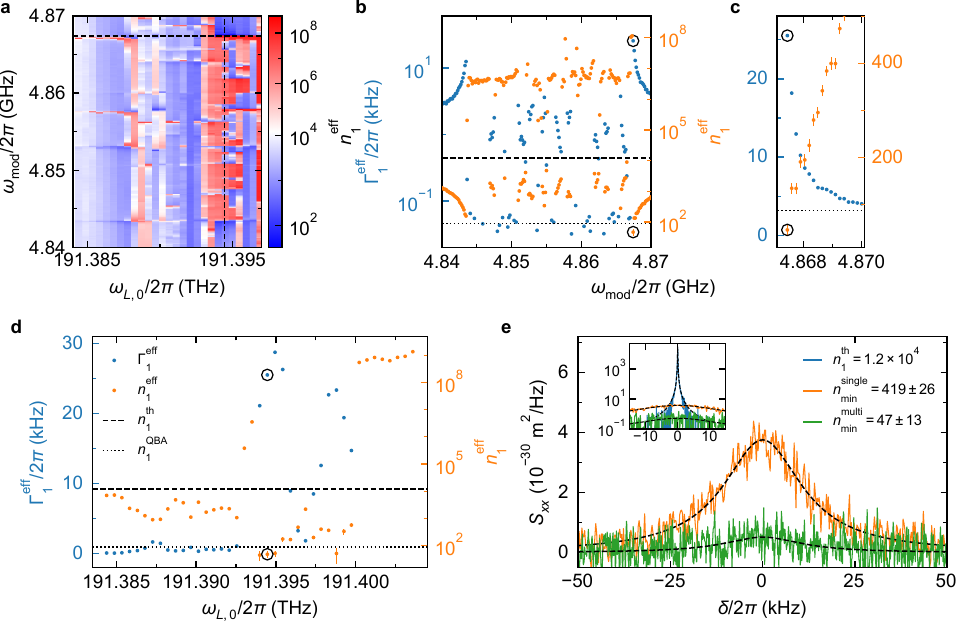}
  \caption{
    \textbf{At higher power the multimode interaction cools the megahertz mode below its single-mode quantum backaction limit.}
    \subfigid{a} The occupancy of mode 1, as a function of the frequency of the central laser tone and the modulation frequency, using an input power of $14.0\uW$ and a modulation depth of $0.45\pi$. Dashed lines mark the slices of \subfigid{b}--\subfigid{d}.
    \subfigid{b}, \subfigid{c} Linewidth (blue) and occupancy (orange) as a function of modulation frequency at constant central laser tone frequency $\omega_{L,0}/2\pi=191.3945$ THz; \subfigid{c} magnifies the region above the upper sum frequency.
    \subfigid{d} Linewidth (blue) and occupancy (orange) as a function of central laser tone frequency at modulation frequency $\wmod/2\pi=4.86745$ GHz. Above $191.397$ THz, which \subfigid{a} leaves out, self-oscillation broadens the probe resonance and strongly changes the slope that sets the transduction (\suppfig{fig:SI:probe-monitor}). In \subfigid{b}--\subfigid{d}, dashed and dotted lines mark $\nth{1}$ and $\nqba{1}$, and circles mark the point with $\neff{1} = 47 \pm 13$.
    \subfigid{e} Spectra of thermal (blue), single-tone backaction-cooled (orange) and multitone backaction-cooled (green, from the circled data points in \subfigid{b}--\subfigid{d}) mechanical motion. The noise floor has been subtracted, as it was significantly different, due to varying setup efficiency.
    Error bars and the uncertainties in \subfigid{e} are one standard deviation, propagated from the uncertainties of the integrated areas of the measured and the reference spectrum.
    \label{fig:res-exp:lw-n1eff-approach_spectra}
  }
\end{figure*}

\Cref{fig:res-exp:freq-lw-n1eff-mainsweep}b, c, e, and f establish that the sign of the backaction is set by $\wmod$, but the occupancies reached at $4.7\uW$ do not yet significantly surpass the single-mode QBA limit of mode 1, $\nqba{1}=\kappa/4\Omega_1=88.9$. We therefore increase the input power to $14.0\uW$ and repeat the experiment, yielding \cref{fig:res-exp:lw-n1eff-approach_spectra}a. The same switching between heating and cooling is observed, though the pattern is more fragmented on the blue-detuned side, where backaction-induced shifts of the mechanical resonance frequencies move the interaction into and out of resonance. A slice with varying modulation frequency at constant central laser tone frequency $\omega_{L,0}/2\pi=191.3945$ THz yields \cref{fig:res-exp:lw-n1eff-approach_spectra}b, in which the linewidth and occupancy rise and fall repeatedly as the interaction sweeps through resonance. Magnifying the region just above the upper sum frequency gives \cref{fig:res-exp:lw-n1eff-approach_spectra}c. Here the occupancy reaches $\neff{1}=47 \pm 13$ phonons immediately before the sum frequency is crossed, significantly below the single-mode QBA limit indicated by the dotted line. The uncertainty is the standard deviation obtained by propagating the uncertainties in the measured and calibration spectrum areas. Since this particular sweep contains only a single point below the single-mode QBA limit, we also plot a slice at constant modulation frequency $\wmod/2\pi=4.86745$ GHz, shown in \cref{fig:res-exp:lw-n1eff-approach_spectra}d. At three neighbouring carrier detunings the occupancy falls to about half the QBA limit at the same modulation frequency. This is the signature expected of the multimode interaction rather than of any detuning-dependent single-mode effect.

The data points that are circled in \cref{fig:res-exp:lw-n1eff-approach_spectra}b-d are obtained from the green spectrum in \cref{fig:res-exp:lw-n1eff-approach_spectra}e. It is plotted there together with the lowest-occupancy spectrum that single-tone dynamical backaction reaches at the same intracavity photon number, taken from a single-tone sweep at $9.4\uW$, and with the unperturbed thermal motion in the inset. Comparing the two, the multimode interaction lowers the experimentally achievable occupancy by almost a factor of ten at equal photon number, determined for both sweeps as described in Methods and compared in \suppsec{subsubsec:SI:ncav} and \suppfig{fig:SI:ncav}. As $\wmod$ lies above $\Omega_1 + \Omega_2$, this operating point is on the sum-frequency branch, whose floor follows the auxiliary occupancy without the Stokes residual of the difference-frequency branch, as shown in \suppsec{subsec:SI:coolinglimit-sum}. At this point mode 1 is damped to $\Geff{1}/2\pi \approx 25$ kHz, more than two orders of magnitude above its intrinsic $\gamma_1/2\pi$.

The single-tone experiment reaches an occupancy of $419\pm26$ phonons. This is above the QBA limit of 88.9 phonons. Its occupancy is set by an excess phonon inflow that scales with the optical damping at both input powers we measured, consistent with a classical force from fluctuations of the cavity frequency relative to the laser, transduced on the flank of the optical resonance. We discuss this effect further in \suppsec{subsec:SI:flux} and \suppfig{fig:SI:single-tone-series}. The multimode damping does not add to this channel: at a fixed carrier detuning the inflow does not follow the damping as $\wmod$ is varied.

\section{Discussion}
A quantitative comparison of the measured operating point with our radiation-pressure model, which uses independently acquired device parameters, is limited by the carrier detuning: the carrier lies close to the heated cavity resonance, where the predicted occupancy depends steeply on the detuning, and the systematic uncertainty of its calibration, described in Methods, is about one cavity linewidth. Across that range the model predicts between about $30$ and $3000$ phonons: at our best estimate of the detuning it predicts far weaker damping and cooling than we observe, while at the red end of the range it reaches the measured occupancy. Dynamical backaction mediated by photothermal rather than radiation-pressure forces, which the numerical model does not include, may contribute a pre-cooling effect. In \suppsec{sec:SI:photothermal} we include this interaction in the multimode model. Photothermal backaction acts far more strongly on the linewidth than on the frequency\cite{jansenControllingPhotothermalForces2026}, so it is not captured by our frequency-based calibration of the optomechanical coupling rate, and it would lower the occupancy from which the multimode scheme begins to cool. Importantly, it cannot account for the sub-QBA result on its own: it is a single-mode effect with no dependence on $\wmod$, whereas the occupancy switches between enhanced cooling and gain on crossing the resonance condition at fixed detuning. Theoretically, it has been argued that photothermal backaction could cool below the radiation-pressure QBA on its own\cite{restrepoClassicalQuantumTheory2011}. However, this is precluded by the large absorption that would be needed, leading to an increased thermal bath temperature. An estimate of the optical heating in our device is given in \suppsec{subsec:SI:expm-heating}.

We have demonstrated that Floquet-driven multimode optomechanics provides a flexible handle on dynamical backaction. A modulated optical drive bridges the large frequency mismatch between a sideband-unresolved MHz mechanical mode and a sideband-resolved GHz auxiliary mode. The resulting optically mediated coupling can be steered from enhanced cooling to gain by tuning the modulation frequency alone, at fixed laser detuning. The modulation frequency sets the relative phase accumulated by fluctuations as they cycle through both mechanical susceptibilities, in direct analogy to how optical detuning sets the sign of single-mode backaction. The bandwidth over which the sign reverses is that of the auxiliary mode rather than of the cavity, four orders of magnitude smaller, which is what makes the switching abrupt.

Exploiting this multimode interaction, we cool the MHz mode to $47 \pm 13$ phonons, approximately half its single-mode quantum backaction limit and nearly ten times lower than what single-tone backaction achieves at the same intracavity photon number. The relevant floor is no longer the quantum backaction limit of the target mode but the occupancy of the auxiliary mode, set by its $4.1$ K bath and by the heat it receives from the target, and amounting to about $18$ phonons at $\Delta_0 = -\Omega_2$. Including the thermal load of the target itself, the full model puts the floor there at about $34$ phonons. Reaching the quantum backaction limit of the auxiliary mode requires $\Gopt{2} \gg \gamma_2 \nth{2}$, and suppressing the thermal load of the target requires $\Gmulti \gg \gamma_1 \nth{1}$. Both call for a lower bath temperature, higher input power, or a higher mechanical $Q$ for the GHz mode.

The structure of \cref{eq:multimode:floor} also indicates avenues for further improvement. Coupling the target to several auxiliary modes in parallel only yields a damping-weighted average of their floors, but gain could come from interference between coupling paths. Interference in the noise, where the vacuum fluctuation that heats mode 1 through Stokes scattering also reaches it through mode 2, cancels the residual $A_{+,1}/\Gmulti$; this is the interference of the proposal\cite{ojanenGroundstateCoolingMechanical2014}, and the sum-frequency branch on which we operated already realizes it. Interference in the coupling, i.e. nonreciprocity, separates the cooling rate, set by $|\sigma_{12}\sigma_{21}|$, from the transfer of the auxiliary occupancy, set by $|\sigma_{12}|/|\sigma_{21}|$. While the auxiliary mode is thermally limited, as it is here, and the thermal load on the target is small, it lowers the floor below $\neff{2}$, by about a third for our device parameters, as we show in \suppsec{subsec:SI:nonreciprocity}. Such nonreciprocal coupling has been realized by interfering two optically mediated transfer paths\cite{xuNonreciprocalControlCooling2019}. The radiation-pressure coupling of our scheme already sums over adjacent tone pairs in $\sigma_{12}$ and is mildly nonreciprocal in the favorable direction, but a useful effect needs control over the amplitude of the lower sideband, for instance through combined amplitude and phase modulation or a separate laser for that tone. These results establish a path toward open-loop quantum control of low-frequency mechanical resonators, which in the sideband-unresolved regime have so far reached the ground state through measurement-based feedback\cite{rossiMeasurementbasedQuantumControl2018,guoActivefeedbackQuantumControl2023}. Because the scheme needs no measurement, it is not limited by detection efficiency, and it opens possibilities in quantum transduction, phonon-based signal processing, and the sympathetic preparation of macroscopic mechanical quantum states.

\section{Methods}
\subsection{Experimental setup}
\label{subsec:meth:expm}

\begin{figure*}[tbp]
  \centering
  \includegraphics[
    width=\textwidth
  ]{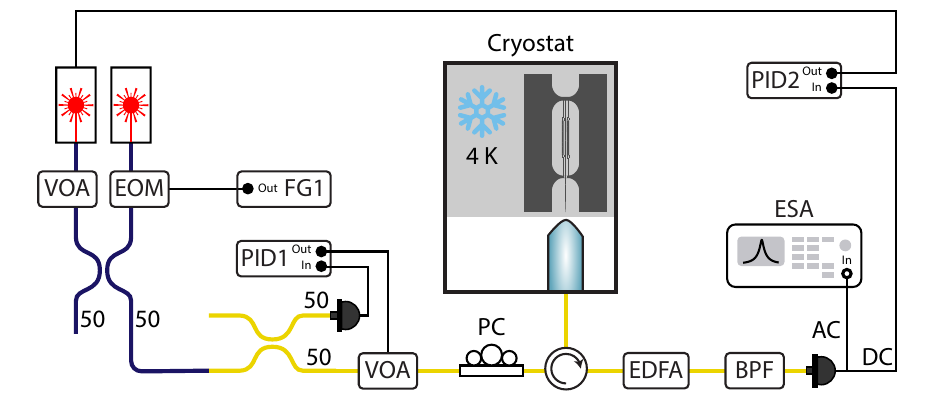}
  \caption{
    Experimental setup. Two tunable lasers are combined, the drive laser being phase-modulated by an EOM driven at $\wmod/2\pi$ and the probe laser strongly attenuated by a VOA. The combined light is coupled to the device in the cryostat through a lensed fiber, and the reflected light is amplified, filtered to isolate the probe, and detected. The AC photocurrent is analyzed on the ESA while the DC component locks the probe laser to the probe optical mode. Component abbreviations are defined in the text.
    \label{fig:meth:exp-setup}
  }
\end{figure*}

The experimental data presented in this work was gathered using the setup depicted in \cref{fig:meth:exp-setup}. Two tunable laser sources (both Toptica CTL 1550) are combined with a 50:50 fiber beamsplitter (Thorlabs PN1550R5A2). One of these lasers is the probe laser. This laser is first attenuated strongly with a variable optical attenuator (Thorlabs VOA50PM-APC; VOA), to around $0.5\uW$ incident on the power meter after the fiber coupler; this is the probe laser. The other laser, which has an input power of a few tens of $\uW$, is called the drive laser and is phase-modulated with frequency $\omega_\mathrm{mod}/2\pi$ by an electro-optic modulator (LN53S-FC; EOM), driven by a function generator (Keysight EXG N5173B; FG1). After the lasers are combined, they pass through another 50:50 beamsplitter, where we use one output to monitor the input power using a power meter (Thorlabs PM100D, S122C head) and another VOA (Thorlabs V1550A) controlled by a PID controller (in-house manufactured PID controller; PID1) to set the monitored input power to a desired value. This input power is multiplied by the measured efficiencies in the setup to calculate the quoted input powers in the main text. The probe-to-drive power ratio is verified at the fiber input before and after each detuning sweep using the power monitor tap to ensure it has not changed.

The combined laser light is then routed through a polarization controller (Thorlabs FPC562; PC) and circulator (Thorlabs 6015-3-APC) into the cryostat, a Montana Cryostation C2. Here, the light is coupled into our device using a lensed fiber. The lensed fiber also collects the reflected light and via the circulator it is passed to an Erbium-doped fiber amplifier (Calmar Coronado AMP-ST-15; EDFA). The light then passes through a tunable band-pass filter (Newport TBF-1550-1.0; BPF) in order to filter out the drive laser, such that only the probe laser is detected on the photodiode (Newport 1544B). The AC component of this detected signal is routed to our real-time electronic spectrum analyzer (Rohde \& Schwarz FSW; ESA), while the DC component is used as the input for another PID controller (RedPitaya 125-14; PID2) to lock the frequency of the probe laser to the probe mode, which can shift due to heating from the drive laser.

\subsection{Device fabrication}
The device was fabricated from a silicon-on-insulator wafer with a $220$~nm silicon device layer on $3~\upmu$m of buried oxide. The pattern was written in hydrogen silsesquioxane resist by electron-beam lithography, developed in tetramethylammonium hydroxide, and transferred into the device layer by inductively coupled plasma reactive-ion etching in a hydrogen bromide and oxygen chemistry. The structures were released by etching the oxide beneath them in hydrofluoric-acid vapor, which also removed the remaining resist.

\subsection{Device parameters}
\label{subsec:meth:dev}

The device, shown in \cref{fig:intro}c, consists of two zipper cavities, each formed by a pair of suspended silicon photonic-phononic crystal nanobeams separated by a narrow slot. A waveguide running between the two cavities couples light in and out.

For this scheme only a single zipper cavity of the device was addressed. Three modes matter for the observed phenomenon. The optical mode used to drive the interaction is the symmetric supermode of the two beams, whose simulated field is shown in \cref{fig:intro}e. The auxiliary GHz mode $\Omega_2$, \cref{fig:intro}d, is localized to the defect region over a few unit cells and is sideband-resolved, $\Omega_2 > \kappa$. The target MHz mode $\Omega_1$, \cref{fig:intro}f, is a flexural mode extending over the full length of the suspended beams and is deeply sideband-unresolved, $\Omega_1 \ll \kappa$.

The measured parameters of these modes, as used in the numerical figures, are listed in \cref{methtab:dev-params}.

\begin{table}[tbp]
  \begin{tabular}{c|c}
    Parameter                     & Value  \\ \hline
    $\Omega_1/2\pi$               & 7.071 MHz \\
    $\gamma_1/2\pi$               & 61 Hz   \\
    $g_1/2\pi$                    & 257 kHz   \\
    $\Omega_{2}/2\pi$             & 4.8506 GHz \\
    $\gamma_{2}/2\pi$             & 89 kHz    \\
    $g_{2}/2\pi$                  & 660 kHz   \\
    $\omega_\mathrm{drive}/2\pi$  & 191.3939 THz  \\
    $\kappa_\mathrm{drive}/2\pi~(\kappa_\mathrm{ex}/2\pi + \kappa_0/2\pi)$  &  2.515 (0.375+2.140) GHz \\
    $\omega_\mathrm{probe}/2\pi$  & 192.8341 THz  \\
    $\kappa_\mathrm{probe}/2\pi~(\kappa_\mathrm{ex}/2\pi + \kappa_0/2\pi)$  & 3.31 (0.519+2.791) GHz  \\
  \end{tabular}
  \caption{Optical and mechanical parameters of the modes used in this work.}
  \label{methtab:dev-params}
\end{table}

\subsection{Occupancy calibration}
\label{subsec:meth:calib-occup}

The frequency $\Omega^{\mathrm{eff}}_1$ and linewidth $\Geff{1}$ follow from a fit of a Lorentzian to each spectrum on a logarithmic scale, with the standard errors of the fit as uncertainties. The occupancy is obtained instead from the integrated area $A$ of the spectrum, converted from dBm per resolution bandwidth to a linear power spectral density, which is proportional to the mean squared displacement and needs no lineshape model, so that it remains valid when the mode self-oscillates. We subtract a constant background, the mean of the lower of the two 200-point windows at the ends of the spectrum, and integrate the remainder with the trapezoidal rule. The uncertainty of $A$ combines the scatter of the background, propagated over the full span, with a term proportional to the signal.

The area is converted to an occupancy with a reference spectrum taken with the drive laser far detuned ($|\Delta_0|/\kappa \gg 10$), where the optical damping is negligible and mode 1 is in equilibrium with the cold finger of the cryostat at $T_0 = 4.1$ K. This temperature is measured by a thermometer next to the sample, which is clamped and thermalized in the same way and shares its radiation environment. Reference spectra are taken immediately before and after each sweep, and with their mean area $A_\mathrm{ref}$ the occupancy is $\neff{1} = \nth{1}(T_0)\,A/A_\mathrm{ref}$. Its relative uncertainty combines those of $A$ and $A_\mathrm{ref}$ as independent errors.

We check three ways in which the reference could differ from the measurement in more than the motion of mode 1. First, the drive may still heat the device at the reference detuning. The reference lies around 20 times further from resonance than the heating maximum (\suppfig[b]{fig:SI:optical-heating}), and since the intensity tails of a Lorentzian fall off as $\Delta^{-2}$, this heating is suppressed by a factor of approximately 400, to less than 0.1 K in the optical mode volume and about 0.004 K in the mechanical bath (\suppsec{subsubsec:SI:mech-bath}). Second, the probe laser heats the device as well. Its input power is 60 times lower than that of the drive laser, and it is locked at a detuning $\Delta_\mathrm{probe}=-\kappa_\mathrm{probe}/2$, which halves its intracavity photon number. This heats the optical mode volume by approximately 0.25 K and, through the same relation, the bath of mode 1 by about 0.01 K. Third, the drive laser is kept on during the reference, so that the gain of the EDFA does not change relative to the measurement; the change of detuning is small compared with the EDFA bandwidth. We neglect all three effects.

The transduction of the probe does change along a sweep. It is proportional to the slope $S$ of the probe reflection at the setpoint of the lock, which we obtain at every drive laser frequency from a reflection scan of the probe mode (\suppfig{fig:SI:probe-monitor}). Since the spectral area scales as $S^2$, we correct every occupancy by the factor $(S_\mathrm{ref}/S)^2$, with $S_\mathrm{ref}$ the mean slope in the two far-detuned reference scans. Away from self-oscillation this correction stays within about $20\%$, comparable to the propagated uncertainties, and at the operating point of \cref{fig:res-exp:lw-n1eff-approach_spectra} it amounts to $1.21$. It reaches up to $8.0$ only in the blue-detuned self-oscillating regime, where the motion broadens the probe resonance and reduces the slope at the setpoint (\suppsec{subsec:SI:probe-monitor}).

\subsection{Drive detuning and intracavity photon number}
\label{subsec:meth:detuning}

Comparing the measurement with the model, and single-tone with multitone cooling at equal intracavity photon number, requires the detuning $\Delta$ of the drive from its heated resonance at every point. The drive mode is not monitored during the sweeps, so we obtain its thermal shift from that of the probe mode\cite{camachoCharacterizationRadiationPressure2009}, $\delta\omega_\mathrm{drive} = R\,\delta\omega_\mathrm{probe}$, with $\delta\omega_\mathrm{probe}$ negative when the device is heated. The drive and probe modes are the even and odd supermodes of the zipper cavity, and equal overlaps with the heated region give equal fractional shifts, so we fix $R = \omega_\mathrm{drive}/\omega_\mathrm{probe} = 0.993$. The cold drive resonance $\omega_c^\mathrm{cold}$ is calibrated with the thermo-optic bistability. In steady state the drive resonance shifts in proportion to the power it absorbs, which is set by its detuning $\Delta = \omega_L - \omega_c^\mathrm{cold} - \delta\omega_\mathrm{drive}$ from the heated resonance\cite{carmonDynamicalThermalBehavior2004, barclayNonlinearResponseSilicon2005, bakerOpticalInstabilitySelfpulsing2012},
\begin{equation}
  \delta\omega_\mathrm{drive} = R\,\delta\omega_\mathrm{probe} = -\delta\omega_\mathrm{max}\,L\big(\omega_L - \omega_c^\mathrm{cold} - \delta\omega_\mathrm{drive}\big),
  \label{eq:meth:bistability}
\end{equation}
with $L(\Delta) = [1 + (2\Delta/\kappa)^2]^{-1}$ the normalized cavity Lorentzian and $\delta\omega_\mathrm{max}$ the shift on resonance. For each laser frequency \cref{eq:meth:bistability} is cubic in $\delta\omega_\mathrm{drive}$, with one or three real roots. The sweep history selects the root, so that the snap of the forward sweep and the release of the backward sweep are outputs of the model to which it can be fitted. We fit the probe shifts of the forward and backward single-tone sweeps at $4.7$ and $9.4\uW$ jointly, with one $\delta\omega_\mathrm{max}$ per power and with the cold drive resonance $\omega_c^\mathrm{cold} = \omega_c^{(0)} + \zeta\,\delta\omega_\mathrm{ref}$ allowed its own response $\zeta$ to the slow drift $\delta\omega_\mathrm{ref}$ of the cold probe resonance. The fitted shifts, $\delta\omega_\mathrm{max}/2\pi = 5.0$ and $10.2$\,GHz, scale with the input power (\suppsec{subsubsec:SI:pump-detuning} and \suppfig{fig:SI:pump-bistability}).

The systematic uncertainty of this calibration comes from $R$ and from the drift of the probe resonance during the heated parts of the sweeps, where it cannot be measured. Refitting with $R$ between $0.89$ and $1.09$, three interpolations of the drift, and either sweep direction places the carrier of the operating point of \cref{fig:res-exp:lw-n1eff-approach_spectra} between $\Delta_0/2\pi = -0.8$ and $2.1$\,GHz, about one cavity linewidth. A Monte Carlo sampling of the same quantities gives the $1\sigma$ interval $1.3$--$2.0$\,GHz.

With the detuning of every point known, the intracavity photon number follows without free parameters. Phase modulation of depth $\beta$ splits the drive into tones $k = -1, 0, +1$ at detunings $\Delta_k = \Delta_0 + k\wmod$ with weights $J_k^2(\beta)$, each filling the cavity according to its own detuning,
\begin{equation}
  n_\mathrm{cav} = \sum_k |\alpha_k|^2 = \frac{4\kex P}{\hbar\omega_L\kappa^2}\,\sum_k J_k^2(\beta)\,L(\Delta_k),
  \label{eq:meth:ncav}
\end{equation}
with $P$ the input power at the device; the single-tone sweeps use the same expression with one tone. Propagated through the Monte Carlo sampling, the detuning calibration makes $n_\mathrm{cav}$ at the operating point uncertain by about $15\%$. The resulting photon numbers of the single-tone and multitone sweeps are compared in \suppsec{subsubsec:SI:ncav}.

\subsection{Use of large language models}
We used a large language model (Claude Opus 4.7 to 5.5 and Fable 5.1, Anthropic) as an assistant in the following tasks: editing the language and structure of text written by the authors, in the manuscript and the Supplementary Information; checking analytical derivations; writing and revising analysis and plotting code, including the scripts that draw the figures. The authors checked all derivations and reviewed and tested all code. The research questions, experiments, interpretation and conclusions are those of the authors.

\section*{Data availability}
The measurement data underlying all figures of this article and its Supplementary Information, together with the intermediate results of the analysis, are available on Zenodo\cite{jansenDataBackactionControl2026}.

\section*{Code availability}
The analysis and theory code that reproduces all figures from the measurement data is available in the same Zenodo record\cite{jansenDataBackactionControl2026}.

\section*{Funding}
This work is part of the research programme of The Netherlands Organisation for Scientific Research (NWO). It is supported by an NWO Vrij Programma (680-92-18-04) grant. It is also funded by the European Union (ERC Consolidator Grant Q-MEME, Grant Agreement No. 101088055). Views and opinions expressed are however those of the authors only and do not necessarily reflect those of the European Union or the European Research Council Executive Agency. Neither the European Union nor the granting authority can be held responsible for them.

\section*{Author contributions}
R.B. designed the initial version of the device and the experiment. M.H.J. extended the device design and the experiment, fabricated the device, performed the measurements, analysed the data and developed the theory. E.V. conceived and supervised the project, and contributed to the interpretation of the data. M.H.J. wrote the manuscript with input from all authors.

\section*{Competing interests}
R.B. is currently employed by QphoX B.V. The other authors declare no competing interests.

\bibliography{references}

\clearpage
\begin{center}
  {\large\bfseries Supplementary Information}
\end{center}
\setcounter{section}{0}
\setcounter{figure}{0}
\setcounter{table}{0}
\setcounter{equation}{0}
\makeatletter
\renewcommand{\thesection}{\arabic{section}}
\renewcommand{\thesubsection}{\thesection.\arabic{subsection}}
\renewcommand{\p@subsection}{}
\renewcommand{\thesubsubsection}{\thesubsection.\arabic{subsubsection}}
\renewcommand{\p@subsubsection}{}
\renewcommand{\theequation}{S\arabic{equation}}
\renewcommand{\thefigure}{\arabic{figure}}
\renewcommand{\figurename}{Supplementary Figure}
\renewcommand{\thetable}{\arabic{table}}
\renewcommand{\tablename}{Supplementary Table}
\renewcommand{\theHsection}{S\arabic{section}}
\renewcommand{\theHsubsection}{S\thesubsection}
\renewcommand{\theHsubsubsection}{S\thesubsubsection}
\renewcommand{\theHequation}{S\arabic{equation}}
\renewcommand{\theHfigure}{S\arabic{figure}}
\renewcommand{\theHtable}{S\arabic{table}}
\def\@sectioncntformat#1{Supplementary Note~\thesection:\quad}
\def\@hangfrom@section#1#2#3{\@hangfrom{#1#2}#3}
\def\@hangfroms@section#1#2{#1#2}
\makeatother
\crefname{section}{Supplementary Note}{Supplementary Notes}
\Crefname{section}{Supplementary Note}{Supplementary Notes}
\crefname{subsection}{Supplementary Note}{Supplementary Notes}
\Crefname{subsection}{Supplementary Note}{Supplementary Notes}
\crefname{subsubsection}{Supplementary Note}{Supplementary Notes}
\Crefname{subsubsection}{Supplementary Note}{Supplementary Notes}
\crefname{figure}{Supplementary Fig.}{Supplementary Figs.}
\Crefname{figure}{Supplementary Figure}{Supplementary Figures}
\crefname{table}{Supplementary Table}{Supplementary Tables}
\Crefname{table}{Supplementary Table}{Supplementary Tables}

\section{Multimode model}
\label{sec:SI:multimode}

This section derives the two-mode description of the main text from the linearized equations of motion of the three-tone drive: the resonant intermode couplings on the difference- and sum-frequency branches, and the diffusion matrix of the optical vacuum noise that enters the Lyapunov equation. The single-mode theory that it extends is summarized in the main text.

\subsection{Linearized equations of motion}
\label{subsec:SI:linearisation}

To study the multimode system, we follow the approach of Ref.~\cite{jiangNonreciprocalDynamicsCryogenic2020}, starting from the equations of motion of two mechanical resonators coupled to a common optical mode via radiation pressure. The cavity has total linewidth $\kappa = \kappa_\mathrm{ex} + \kappa_0$, where $\kappa_\mathrm{ex}$ is the external coupling rate with vacuum noise input $\hat{a}_\mathrm{in}$ and $\kappa_0$ is the intrinsic loss rate with an independent vacuum noise input $\hat{a}_{\mathrm{in},0}$\cite{chanLaserCoolingOptomechanical2012}:

\begin{subequations}
  \begin{align}
    \dot{\hat{a}} &= -\left(\frac{\kappa}{2} + i\omega_c\right)\hat{a}
    - i\sum_{n=1}^{2} g_n\left(\hat{b}_n + \hat{b}_n^\dagger\right)\hat{a} \nonumber\\
    &\quad + \sqrt{\kex}\,\hat{a}_\mathrm{in}
    + \sqrt{\kint}\,\hat{a}_{\mathrm{in},0}, \\
    \dot{\hat{b}}_n &= -\left(\frac{\gamma_n}{2} + i\Omega_n\right)\hat{b}_n
    - ig_n\hat{a}^\dagger\hat{a} + \sqrt{\gamma_n}\,\hat{\eta}_n,
    \quad n \in \{1,2\}.
  \end{align}
\end{subequations}

The input field is modulated at frequency $\wmod$ with modulation depth $\beta$, producing three tones in the cavity coherent amplitude. We therefore substitute $\hat{a} = (\alpha(t) + \hat{d}(t))e^{-i\omega_c t}$, with

\begin{equation}
  \alpha(t) = \sum_{k=-1}^{+1} \alpha_k\,e^{-i(\Delta_k t + \phi_k)}, \qquad
  \alpha_k = \frac{\sqrt{\kex}\,\alpha_{\mathrm{in},k}}{\kappa/2 - i\Delta_k},
\end{equation}
where $\Delta_k = \Delta_0 + k\,\wmod$ are the detunings of the three tones, $\phi_k$ their phases, and $\alpha_{\mathrm{in},k} \propto J_k(\beta)$ the input amplitudes, with $J_k$ the $k$-th Bessel function of the first kind. The sum is truncated after the first-order sidebands, which carry about $90\%$ of the power at the modulation depths used here. Including the second order ($J_2(0.45\pi) \approx 0.21$) changes the lowest occupancy, reached with the carrier close to the cavity resonance, by less than $1\%$. It does matter at $\Delta_0 = -\Omega_2$ on the sum-frequency branch, where the $k = +2$ tone falls on the blue sideband of mode 2 and heats it; the occupancies quoted at that reference point are first-order values. For pure phase modulation the tone phases are $\phi_k = k\phi_\mathrm{mod}$, so the phase differences $\phi_l - \phi_k$ that appear below depend only on $l-k$; a global shift of $\phi_\mathrm{mod}$ can be absorbed into the phase reference of the mechanical operators and is therefore dropped from here on. It is retained explicitly through \cref{eq:SI:fullsum} because unequal tone phases are what enables nonreciprocal inter-mode coupling (\cref{subsec:SI:nonreciprocity}). Absorbing the static radiation-pressure displacement into each $\hat{b}_n$ and retaining only terms linear in the fluctuation operator $\hat{d}$, the linearized equations of motion are

\begin{subequations}
  \begin{align}
    \dot{\hat{d}}(t) &= -\frac{\kappa}{2}\,\hat{d}(t)
    - i\sum_{n=1}^{2} g_n\left(\hat{b}_n(t) + \hat{b}_n^\dagger(t)\right)\alpha(t) \nonumber\\
    &\quad + \sqrt{\kex}\,\hat{a}_\mathrm{in}(t)
    + \sqrt{\kint}\,\hat{a}_{\mathrm{in},0}(t), \\
    \dot{\hat{b}}_n(t) &= -\left(\frac{\gamma_n}{2} + i\Omega_n\right)\hat{b}_n(t) \nonumber\\
    &\quad - ig_n\left(\alpha^*(t)\hat{d}(t) + \alpha(t)\hat{d}^\dagger(t)\right)
    + \sqrt{\gamma_n}\,\hat{\eta}_n(t).
  \end{align}
\end{subequations}

Transforming to the Fourier domain, $\hat{O}[\omega] = \int \hat{O}(t)\,e^{i\omega t}\,dt$, and using the conjugation property $\hat{d}^\dagger[\omega] = (\hat{d}[-\omega])^\dagger$, the optical fluctuation operator and mechanical sidebands are

\begin{subequations}
  \begin{align}
    \hat{d}[\omega] &= -i\chi(\omega)\sum_{n=1}^{2}\sum_{k=-1}^{+1}
    g_n\,\alpha_k\,e^{-i\phi_k} \nonumber\\
    &\quad\times\left(\hat{b}_n[\omega - \Delta_k] + \hat{b}_n^\dagger[\omega - \Delta_k]\right), \\
    \hat{b}_n[\omega] &= -i\chi_n(\omega)\,g_n\Bigl(
      \sum_{k=-1}^{+1}\alpha_k^*\,e^{i\phi_k}\,\hat{d}[\omega + \Delta_k] \nonumber\\
      &\qquad + \alpha_k\,e^{-i\phi_k}\,\hat{d}^\dagger[\omega - \Delta_k]
    \Bigr),
  \end{align}
\end{subequations}
where $\chi(\omega) = (\kappa/2 - i\omega)^{-1}$ and $\chi_n(\omega) = (\gamma_n/2 - i(\omega - \Omega_n))^{-1}$. Substituting the expression for $\hat{d}[\omega]$ (and its conjugate) into the mechanical equation gives

\begin{align}
  \hat{b}_n[\omega] ={}& \chi_n(\omega)\,g_n \sum_{k,l=-1}^{+1}\sum_{p=1}^{2}
  \alpha_l^*\,\alpha_k\,e^{i(\phi_l - \phi_k)}\,g_p \nonumber\\
  &\times\left[\chi(\omega - \Delta_k) - \chi(\omega + \Delta_l)\right] \nonumber\\
  &\times\left(\hat{b}_p[\omega + \Delta_{lk}] + \hat{b}_p^\dagger[\omega + \Delta_{lk}]\right),
  \label{eq:SI:fullsum}
\end{align}
where $\Delta_{lk} = \Delta_l - \Delta_k = (l-k)\,\wmod$ is the frequency difference between tones $l$ and $k$. The double tone sum runs over $9$ pairs $(k,l)$ and the mode sum over $p \in \{1,2\}$, giving $18$ terms, each carrying both a $\hat{b}_p$ and a $\hat{b}_p^\dagger$ contribution.

We simplify this by exploiting the high mechanical $Q$-factor ($\gamma_p \ll \Omega_p$). The mechanical susceptibility $\chi_n(\omega)$ on the left is sharply peaked at $\omega = \Omega_n$, so a term contributes appreciably only if the operator on the right is also evaluated near one of its own resonances, that is if $\Omega_n + \Delta_{lk} \approx \pm\Omega_p$ to within $\gamma_p$. Because $\Delta_{lk}$ can only take the values $0, \pm\wmod, \pm2\wmod$, this selects a small subset:

\begin{itemize}
  \item $\Delta_{lk} = 0$ (three pairs, $l = k$) with $p = n$. These are the diagonal, single-tone-like terms and give $\sigma_{nn}$. They are resonant for any $\wmod$.
  \item $\Delta_{lk} = \pm\wmod$ (four pairs, $l = k \pm 1$) with $p \neq n$. These are resonant only when $\wmod \approx \Omega_2 - \Omega_1$, in which case they couple $\hat{b}_1 \leftrightarrow \hat{b}_2$ through a beamsplitter interaction and give $\sigma_{12}$, $\sigma_{21}$; or when $\wmod \approx \Omega_2 + \Omega_1$, in which case they couple $\hat{b}_1 \leftrightarrow \hat{b}_2^\dagger$ through a two-mode squeezing interaction and give the sum-frequency couplings of \cref{subsec:SI:sumfreq}.
  \item $\Delta_{lk} = \pm 2\wmod$ (two pairs, $l = k \pm 2$). These would require $2\wmod \approx \Omega_2 \pm \Omega_1$ and are off-resonant everywhere in the parameter range studied here.
\end{itemize}

All remaining terms are off resonance by at least $2\Omega_1$, and are suppressed by a factor of order $\gamma_p/2\Omega_1$, or $|\sigma|/2\Omega_1$ for the optically broadened mode 2, at most about $1\%$ for $\Gtot{2}/2\pi \lesssim 150$ kHz; they are therefore dropped, which constitutes the rotating wave approximation in the mechanical frame. Retaining the resonant terms and transforming back to the time domain at $\omega = \Omega_n$ yields the equations of motion and the self-energies $\sigma_{nn}$, $\sigma_{12}$, $\sigma_{21}$ of the main text, equations~(\ref{eq:multimode:EOM}) and~(\ref{eq:multimode:sigmas}). The tone phases enter only through $e^{i(\phi_l-\phi_k)} = e^{\pm i\phi_\mathrm{mod}}$ for $l = k\pm1$; these opposite phases attach to $\sigma_{12}$ and $\sigma_{21}$ respectively, but cancel in the product $\sigma_{12}\sigma_{21}$.

\subsection{Diffusion matrix and Lyapunov equation}
\label{subsec:SI:Q}

In practice, $Q$ combines contributions from the mechanical thermal baths and the optical quantum backaction. Both mechanical modes are driven by the same intracavity field, so the optical vacuum fluctuations induce not only per-mode heating but also cross-correlated backaction between the two modes. The full diffusion matrix is therefore

\begin{equation}
  Q =
  \begin{pmatrix}
    \gamma_1 \nth{1} + A_{+,1} & A_{\times,12} \\
    \left(A_{\times,12}\right)^* & \gamma_2 \nth{2} + A_{+,2}
  \end{pmatrix}.
  \label{eq:SI:Qfull}
\end{equation}

The thermal contributions $\gamma_n \nth{n}$ are strictly diagonal because the two mechanical baths are independent by assumption. The optical contributions originate from a common source and can be derived from the noise correlators.

After adiabatic elimination of the optical field, the optical noise force on mode $n$ in the Fourier domain receives contributions from both the external coupling port and the intrinsic loss port;

\begin{equation}
  \begin{split}
    \hat{F}_n[\omega] = -ig_n\sum_k\Bigl(
      & \alpha_k^*\,\chi(\omega+\Delta_k)\,\bigl(
        \sqrt{\kex}\,\hat{a}_\mathrm{in}[\omega+\Delta_k] \\
      & \qquad + \sqrt{\kint}\,\hat{a}_{\mathrm{in},0}[\omega+\Delta_k]
      \bigr) \\
      & + \alpha_k\,\chi(\omega-\Delta_k)\,\bigl(
        \sqrt{\kex}\,\hat{a}_\mathrm{in}^\dagger[\omega-\Delta_k] \\
      & \qquad + \sqrt{\kint}\,\hat{a}_{\mathrm{in},0}^\dagger[\omega-\Delta_k]
      \bigr)
    \Bigr),
  \end{split}
\end{equation}
where $\chi(\omega) = (\kappa/2 - i\omega)^{-1}$ and the mechanical backaction terms have been dropped as they do not contribute to the noise correlator. With several drive tones the optical noise force is not stationary, and the diffusion matrix must be defined in the co-rotating frame in which the Lyapunov equation is written. We first evaluate the laboratory-frame correlator. Using the vacuum input noise correlators, which in this convention follow from $\langle\hat{a}_\mathrm{in}(t)\hat{a}_\mathrm{in}^\dagger(t')\rangle = \delta(t-t')$ as $\langle\hat{a}_\mathrm{in}[\omega]\hat{a}_\mathrm{in}^\dagger[\omega']\rangle = 2\pi\,\delta(\omega+\omega')$, $\langle\hat{a}_{\mathrm{in},0}[\omega]\hat{a}_{\mathrm{in},0}^\dagger[\omega']\rangle = 2\pi\,\delta(\omega+\omega')$, and $\langle\hat{a}_\mathrm{in}^\dagger\hat{a}_\mathrm{in}\rangle = \langle\hat{a}_{\mathrm{in},0}^\dagger\hat{a}_{\mathrm{in},0}\rangle = 0$, the two vacuum ports are statistically independent so their noise contributions add incoherently, yielding a combined prefactor $\kex + \kint = \kappa$. The normal-ordered force correlator is then

\begin{equation}
  \begin{split}
  \langle\hat{F}_m^\dagger[\omega']\,\hat{F}_n[\omega]\rangle
  &= 2\pi\, g_n g_m\,\kappa \sum_{k,l} \alpha_k\,\alpha_l^*\\
  &\quad\times\chi(\omega-\Delta_k)\,\chi^*(-\omega'-\Delta_l)\\
  &\quad\times\delta\bigl(\omega+\omega' + (l-k)\,\wmod\bigr).
  \end{split}
  \label{eq:SI:Qcorrelator}
\end{equation}

The same-tone terms $k = l$ carry $\delta(\omega+\omega')$ and are stationary. The adjacent-tone terms $l = k\pm1$ carry $\delta(\omega+\omega'\pm\wmod)$: they describe a single vacuum fluctuation that is a sideband of tone $k$ for one mode and of tone $k\pm1$ for the other, and their statistics oscillate at the modulation frequency rather than being time-independent. In the co-rotating frame $\bm{b}^r = U\bm{b}$ defined in \maineqref{eq:multimode:Drot} the $(1,2)$ element of the noise correlator acquires a factor $e^{-i\wmod(t+t')/2}$, which renders its adjacent-tone component stationary and makes its same-tone component oscillate at $\wmod$; the latter averages out under the same rotating wave approximation that renders $D^\mathrm{rot}$ time-independent, while the diagonal elements retain their stationary same-tone component. Equivalently, $Q_{nm}$ is the coefficient of $2\pi\,\delta$ in \cref{eq:SI:Qcorrelator}, obtained by evaluating the noise on mode $n$ at $\omega = \Omega_n$ and that on mode $m$ at $\omega' = -\Omega_m$, and keeping the term whose delta function is satisfied there: $l = k$ for $n = m$, and $l = k+1$ for $(n,m) = (1,2)$ at $\wmod = \Omega_2 - \Omega_1$.

For $n = m$, using $|\chi(\omega)|^2 = [(\kappa/2)^2 + \omega^2]^{-1}$, this gives the per-mode optical heating rate

\begin{equation}
  \label{eq:SI:Aplus}
  A_{+,n} = g_n^2\,\sum_{k=-1}^{+1}|\alpha_k|^2 \frac{\kappa}{(\kappa/2)^2 + (\Delta_k - \Omega_n)^2},
\end{equation}
which is \maineqref{eq:multimode:Aplus} and reduces, for a single tone, to the Stokes rate $A_+ = |g_0\alpha|^2\kappa\,[(\kappa/2)^2 + (\Delta-\Omega_m)^2]^{-1}$ that enters equation~(\ref{eq:Gammaopt}) there. For $n \neq m$ the stationary rotating-frame cross-diffusion near the difference frequency is

\begin{equation}
  \label{eq:SI:Across}
  A_{\times,12} = g_1 g_2\,\kappa \sum_{k=-1}^{0} \alpha_k\,\alpha_{k+1}^*\,
  \chi(\Omega_1 - \Delta_k)\,\chi^*(\Omega_2 - \Delta_{k+1}),
\end{equation}
with $Q_{21} = (A_{\times,12})^*$ by Hermiticity of $Q$, which also follows from \cref{eq:SI:Qcorrelator} with $l = k-1$. At the resonance $\wmod = \Omega_2 - \Omega_1$ the two susceptibilities coincide, $\Omega_2 - \Delta_{k+1} = \Omega_1 - \Delta_k$, and each term reduces to $\alpha_k\alpha_{k+1}^*|\chi(\Omega_1-\Delta_k)|^2$.

The magnitude of the cross-diffusion is set by the cavity filtering of the shared vacuum photon. Near the difference frequency with red-detuned tones, the shared photon is the Stokes sideband of both modes: for the pair $\Delta_k = -\Omega_2$, $\Delta_{k+1} = -\Omega_1$ it sits at $-(\Omega_1+\Omega_2)$ from cavity resonance, and since $\Omega_1 \ll \Omega_2$, \cref{eq:SI:Across} is suppressed relative to $A_{+,1}$ by

\begin{equation}
  \epsilon \equiv \frac{|\chi(\Omega_2)|^2}{|\chi(0)|^2}
  = \frac{(\kappa/2)^2}{(\kappa/2)^2 + \Omega_2^2} \approx 0.063,
  \label{eq:SI:epscross}
\end{equation}
as in \maineqref{eq:multimode:epscross}. On the sum-frequency branch the situation is reversed, as we show in \cref{subsec:SI:sumfreq}.

To confirm consistency with the single-mode result, we take the limit of a single tone and no inter-mode coupling, so that $\sigma_{12} = \sigma_{21} = 0$, $A_{\times,12} = 0$, and the Lyapunov equation decouples. The $(1,1)$ element gives

\begin{equation}
  (D_{11}+D^*_{11}) V_{11} + Q_{11} = 0 \quad \Rightarrow \quad \neff{1} = \frac{\gamma_1 \nth{1} + A_{+,1}}
  {\gamma_1 + \Gopt{1}},
\end{equation}
where $\Gopt{1} = -2\,\mathrm{Re}\{D_{11}\} - \gamma_1 = A_{-,1} - A_{+,1}$. In the limit of strong optomechanical damping ($\Gopt{1} \gg \gamma_1$), this reduces to:

\begin{equation}
  \neff{1} \to \frac{A_{+,1}}{A_{-,1} - A_{+,1}}
  = \nqbasm = \frac{\kappa}{4\Omega_1},
\end{equation}
where the final equality follows from substituting the Lorentzian forms of $A_{\pm,1}$ at the optimal detuning of the sideband-unresolved regime, $\Delta_0 = -\kappa/2$, recovering \maineqref{eq:nqba-unresolved}.

The steady-state Lyapunov equation $D^\mathrm{rot}V + V(D^\mathrm{rot})^\dagger + Q = 0$ is linear in the four elements of $V$. With the row-major vectorization $\bm{v} = (V_{11}, V_{12}, V_{21}, V_{22})^T$, for which $\mathrm{vec}(MVN^T) = (M\otimes N)\,\mathrm{vec}(V)$, it becomes

\begin{equation}
  \left(D^\mathrm{rot} \otimes \mathbf{1} + \mathbf{1} \otimes D^{\mathrm{rot}*}\right)\bm{v} = -\mathrm{vec}(Q),
  \label{eq:SI:lyapunovsuperop}
\end{equation}
where the eigenvalues of the $4\times4$ matrix on the left are the pairwise sums $\lambda_i + \lambda_j^*$ of the eigenvalues of $D^\mathrm{rot}$. It is therefore invertible, and the steady state unique, whenever both eigenvalues of $D^\mathrm{rot}$ have negative real parts. Grid points at which either eigenvalue has a positive real part are dynamically unstable and are masked in all figures. On the grids of carrier detuning $\Delta_0$ and modulation frequency $\wmod$ of the main text, \cref{eq:SI:lyapunovsuperop} is solved at all grid points at once by a batched linear solve, and the occupancies are $\neff{1} = V_{11}$ and $\neff{2} = V_{22}$.

\subsection{Sum-frequency interaction}
\label{subsec:SI:sumfreq}

When $\wmod \approx \Omega_2 + \Omega_1$, the resonant terms in \cref{eq:SI:fullsum} are again those with $|\Delta_{lk}| = \wmod$, but they now couple $\hat{b}_1$ to $\hat{b}_2^\dagger$ rather than to $\hat{b}_2$. In the interaction picture $\hat{b}_1 \propto e^{-i\Omega_1 t}$ while $\hat{b}_2^\dagger \propto e^{+i\Omega_2 t}$, so the coupling term must carry $e^{-i\wmod t}$ for the susceptibilities to be resonant on $\wmod = \Omega_1 + \Omega_2$. This is the opposite sense to the difference-frequency case, \maineqref{eq:multimode:EOM}, and the equations of motion become

\begin{align}
  \label{eq:SI:sumfreqEOM}
  \begin{split}
    \dot{\hat{b}}_1(t) &= -\left(\frac{\gamma_1}{2} + i\Omega_1 + \sigma_{11}\right)\hat{b}_1(t)
    - \sigma_{12}^s\,e^{-i\wmod t}\,\hat{b}_2^\dagger(t) \\
    &\quad + \sqrt{\gamma_1}\,\hat{\eta}_1, \\
    \dot{\hat{b}}_2^\dagger(t) &= -\left(\frac{\gamma_2}{2} - i\Omega_2
    + \sigma_{22}^*\right)\hat{b}_2^\dagger(t)
    - \sigma_{21}^s e^{+i\wmod t}\,\hat{b}_1(t) \\
    &\quad + \sqrt{\gamma_2}\,\hat{\eta}_2^\dagger.
  \end{split}
\end{align}
The sum-frequency self-energies are:

\begin{subequations}
  \begin{align}
    \sigma_{nn} &= g_n^2\sum_{k=-1}^{+1}|\alpha_k|^2
    \left(\chi(\Omega_n + \Delta_k) - \chi(\Omega_n - \Delta_k)\right), \\
    \sigma_{12}^s &= g_1 g_2\sum_{k=-1}^{0}\alpha_{k}^*\,\alpha_{k+1}
    \left(\chi(\Omega_1 + \Delta_{k}) - \chi(\Omega_1 - \Delta_{k+1})\right), \\
    \sigma_{21}^s &= g_2 g_1\sum_{k=-1}^{0}\alpha_{k+1}^*\,\alpha_{k} \nonumber\\
    &\quad\times\left(\chi(-\Omega_2 - \Delta_{k}) - \chi(-\Omega_2 + \Delta_{k+1})\right).
  \end{align}
  \label{eq:SI:sumfreqsigmas}
\end{subequations}

The diagonal self-energy is unchanged from the difference-frequency case, \maineqref{eq:multimode:sigmas}, and carries no superscript, because it depends only on the individual tone powers and not on inter-tone sideband interference. The off-diagonal terms $\sigma_{12}^s$ and $\sigma_{21}^s$ differ from their difference-frequency counterparts in the ordering of conjugation and, for $\sigma_{21}^s$, in the sign of the $\chi$ arguments.

To construct the time-independent dynamical matrix we write $\bm{b}^s = (\hat{b}_1, \hat{b}_2^\dagger)^T$ and move to a co-rotating frame via $\bm{b}^{s,r}(t) = U^s(t)\bm{b}^s(t)$. Because the coupling exponentials have the opposite sign to the difference-frequency case, the required unitary is also reversed relative to the $U$ in \maineqref{eq:multimode:Drot}:

\begin{equation}
  U^s(t) = \mathrm{diag}\!\left(e^{+i\wmod t/2},\; e^{-i\wmod t/2}\right),
  \label{eq:SI:Usumfreq}
\end{equation}
giving:

\begin{align}
  \begin{split}
  D^{\mathrm{rot},s} ={}& -i\begin{pmatrix}
    \Omega_1 - \wmod/2 & 0 \\
    0 & -\Omega_2 + \wmod/2
  \end{pmatrix}\\
  &- \begin{pmatrix}
    \gamma_1/2 + \sigma_{11} & \sigma_{12}^s \\
    \sigma_{21}^s & \gamma_2/2 + \sigma_{22}^*
  \end{pmatrix}.
  \end{split}
  \label{eq:SI:Drotsumfreq}
\end{align}

The two diagonal entries become degenerate at $\wmod = \Omega_1 + \Omega_2$, as required. The key structural difference from the difference-frequency matrix \maineqref{eq:multimode:Drot} is the conjugation of the $i\Omega_2$ and $\sigma_{22}$ terms in the $(2,2)$ entry, which arises because the dynamical variable is now $\hat{b}_2^\dagger$ rather than $\hat{b}_2$. As a consequence the $(2,2)$ entry has a positive imaginary part in the rotating frame, and the interaction is of two-mode-squeezing rather than beam-splitter type: the interaction creates and annihilates phonon pairs rather than exchanging them, so the cooling and anti-damping lobes are mirrored relative to the difference-frequency case.

The Lyapunov equation $D^{\mathrm{rot},s} V^s + V^s (D^{\mathrm{rot},s})^\dagger + Q^s = 0$ is solved for the covariance matrix $V^s_{nm} = \langle (\bm{b}^s_m)^\dagger \bm{b}^s_n\rangle$, so that $V^s_{22} = \langle\hat{b}_2\hat{b}_2^\dagger\rangle = \neff{2} + 1$ rather than $\neff{2}$. The diffusion matrix must be conjugated in the same way. Its optical term on mode 2 becomes the anti-Stokes rate $A_{-,2}$, obtained from \cref{eq:SI:Aplus} with $\Omega_2 \to -\Omega_2$, and its mechanical term becomes $\gamma_2(\nth{2}+1)$, because $\langle\hat{b}_2\hat{b}_2^\dagger\rangle$ relaxes to $\nth{2}+1$ under bare damping:

\begin{equation}
  Q_{22}^s = \gamma_2\left(\nth{2}+1\right) + A_{-,2}.
  \label{eq:SI:Q22sum}
\end{equation}
Under the carrier tone alone this returns $V_{22}^s = Q_{22}^s/\Gtot{2} = \neff{2}+1$ exactly, using $A_{-,2} - A_{+,2} = \Gopt{2}$, whereas retaining $\gamma_2\nth{2}$ would understate $V_{22}^s$ by $\gamma_2/\Gtot{2}$.

The cross-diffusion element is now the correlator $\langle\hat{F}_2[\omega']\hat{F}_1[\omega]\rangle$ evaluated at $\omega = \Omega_1$, $\omega' = \Omega_2$, which selects $l = k-1$ in \cref{eq:SI:Qcorrelator} and gives

\begin{equation}
  \label{eq:SI:Q12sum}
  A_{\times,12}^s = -\,g_1 g_2\,\kappa \sum_{k=-1}^{0} \alpha_k^*\,\alpha_{k+1}\,
  \chi(\Omega_2 + \Delta_k)\,\chi(\Omega_1 - \Delta_{k+1}),
\end{equation}
which at $\wmod = \Omega_1 + \Omega_2$ likewise collapses to $-\alpha_k^*\alpha_{k+1}|\chi(\Omega_1 - \Delta_{k+1})|^2$ per term. Unlike its difference-frequency counterpart it is not suppressed: the anti-Stokes sideband of tone $k$ on mode 2 and the Stokes sideband of tone $k+1$ on mode 1 are emitted into the same cavity-resonant frequency, and $A_{\times,12}^s$ is of relative size $g_2|\alpha_k|/(g_1|\alpha_{k+1}|)$ compared with $A_{+,1}$. Comparing \cref{eq:SI:Q12sum} with $\sigma_{12}^s$ of \cref{eq:SI:sumfreqsigmas} for a single tone pair with the auxiliary tone on its anti-Stokes resonance, $\Delta_k = -\Omega_2$, gives

\begin{equation}
  A_{\times,12}^s = \frac{2\,\sigma_{12}^s}{1 - (\kappa/2)\,\chi(\Omega_1-\Omega_2)} \approx 2\,\sigma_{12}^s,
  \label{eq:SI:Q12sum_sigma}
\end{equation}
where the denominator, $0.94 + 0.24\,i$ for our parameters, stems from the off-resonant Stokes path in $\sigma_{12}^s$, of relative amplitude $|(\kappa/2)\,\chi(\Omega_1-\Omega_2)| \approx \sqrt{\epsilon} \approx 0.25$.

This is not a coincidence of the single-pair geometry. The cross-diffusion that enters the Lyapunov equation is the ordered correlator $\langle\hat{F}_2\hat{F}_1\rangle$, and exactly at $\wmod = \Omega_1 + \Omega_2$, for any set of tone amplitudes and phases, the two orderings of the cross noise differ by the coupling,

\begin{equation}
  \langle\hat{F}_2\hat{F}_1\rangle - \langle\hat{F}_1\hat{F}_2\rangle
  = \sigma_{12}^s - \sigma_{21}^{s*},
  \label{eq:SI:noisecommutator}
\end{equation}
which follows from \cref{eq:SI:sumfreqsigmas} with $\chi(-\omega) = \chi^*(\omega)$ and $2\,\mathrm{Re}\,\chi = \kappa|\chi|^2$. This is the zero-temperature fluctuation-dissipation relation of the cavity bath, the same ordering difference that separates $A_{+,n}$ from $A_{-,n}$ on the diagonal. The anti-normally ordered correlator $\langle\hat{F}_1\hat{F}_2\rangle$ involves the photon at $\Omega_1 + \Delta_k \approx -\Omega_2$, which is far from the cavity resonance, and is suppressed by the factor $\epsilon$ of \cref{eq:SI:epscross}; dropping it gives $A_{\times,12}^s \approx \sigma_{12}^s - \sigma_{21}^{s*}$, of which \cref{eq:SI:Q12sum_sigma} is the reciprocal single-pair case. This relation is what allows the vacuum noise to interfere on the sum-frequency branch, as we show in \cref{subsec:SI:coolinglimit-sum}.

\section{Cooling limits}
\label{sec:SI:coolinglimits}

We show that the occupancy of mode 1 that the multimode scheme can reach is bounded below by the occupancy of mode 2, up to a residual that differs between the two branches. On the difference-frequency branch, with the carrier on the red sideband of mode 2, the beam-splitter coupling leaves at least of order one phonon of Stokes heating in mode 1. On the sum-frequency branch the vacuum noise interferes through the cross-diffusion of \cref{subsec:SI:sumfreq} and removes this residual, so that the floor follows the auxiliary occupancy. Nonreciprocal coupling, finally, lowers the share of that occupancy that reaches mode 1.

\subsection{Difference-frequency branch}
\label{subsec:SI:coolinglimit}

Starting from the Lyapunov equation $D^{\mathrm{rot}}V + V(D^{\mathrm{rot}})^\dagger + Q = 0$, we extract the $(1,1)$, $(1,2)$, and $(2,2)$ components explicitly. Writing $\Gtot{n} = \gamma_n + \Gopt{n}$ for the total damping rate of mode $n$, with $\Gopt{n} = 2\,\mathrm{Re}\,\sigma_{nn}$ the optomechanical contribution, and using $V_{21} = V_{12}^*$ because of the Hermiticity of $Q$, these are:

\begin{subequations}
  \begin{align}
    -\Gtot{1} V_{11} - 2\,\mathrm{Re}(\sigma_{12}\,V_{12}^*) + Q_{11} &= 0, \label{eq:SI:lyap11}\\
    (D_{11}+D_{22}^*)\,V_{12} - \sigma_{12}\,V_{22} - \sigma_{21}^*\,V_{11} + Q_{12} &= 0, \label{eq:SI:lyap12}\\
    -\Gtot{2} V_{22} - 2\,\mathrm{Re}(\sigma_{21}\,V_{12}) + Q_{22} &= 0. \label{eq:SI:lyap22}
  \end{align}
  \label{eq:SI:lyapcomponents}
\end{subequations}

For the remainder of this analysis, we set the modulation near the difference-frequency resonance, $\wmod \approx \Omega_2 - \Omega_1$, and the carrier detuning to the red sideband of mode 2, $\Delta_0 = -\Omega_2$.

We consider the regime in which mode 2 is strongly cooled by the carrier tone independently of the multimode coupling, $\Gtot{2} \gg \Gtot{1}, |\sigma_{12}|,|\sigma_{21}|$, so that the coupling terms in \cref{eq:SI:lyap22} are a small perturbation and

\begin{equation}
  \neff{2} = V_{22} \approx \frac{Q_{22}}{\Gtot{2}}
  = \frac{\gamma_2\nth{2} + A_{+,2}}{\gamma_2 + \Gopt{2}}
  = n_2^{\mathrm{eff,single}},
  \label{eq:SI:V22approx}
\end{equation}
the occupancy that mode 2 reaches under the carrier tone alone. It reduces to a backaction limit only when $A_{+,2} \gg \gamma_2 \nth{2}$, which is not the case in our device, so we carry $\neff{2}$ through the derivation and take the limit of a backaction-limited auxiliary mode only at the end. That limit is not $\nqba{2}$: the heating rate $A_{+,2}$ in \cref{eq:SI:V22approx} sums over all three tones, and the tone that addresses mode 1 sits close to the cavity resonance, so its Stokes sideband on mode 2 at $\Delta_{k+1} - \Omega_2 \approx -\Omega_2$ is suppressed only by the factor $\epsilon \approx 0.063$ of \cref{eq:SI:epscross} rather than by $\nqba{2} = |\chi(2\Omega_2)|^2/|\chi(0)|^2 \approx 0.017$. Even at zero bath temperature the auxiliary mode therefore retains $\neff{2} \gtrsim \epsilon\,|\alpha_{k+1}|^2/|\alpha_k|^2$. For pure phase modulation at $\Delta_0 = -\Omega_2$, where the cavity filter fixes $|\alpha_{+1}|/|\alpha_0| = |\chi(\Omega_1)|J_1(\beta)/(|\chi(\Omega_2)|J_0(\beta))$, this amounts to $J_1^2(\beta)/J_0^2(\beta)$, about one phonon at $\beta = 0.45\pi$, independent of the input power. We note that \cref{eq:SI:V22approx} also neglects the heat that flows into mode 2 through the coupling; the term $2\,\mathrm{Re}(\sigma_{21}V_{12})$ in \cref{eq:SI:lyap22}. At our bath temperature, this flow (of order $\Gmulti\neff{1}$) is comparable to the thermal load $\gamma_2\nth{2}$, and the auxiliary occupancy at the reference point of the main text is correspondingly higher than the single-tone value. All occupancies quoted in the main text are therefore taken from the full solution of the Lyapunov equation.

Because $\Gtot{2} \gg \Gtot{1}$, the coefficient of $V_{12}$ in \cref{eq:SI:lyap12} is approximately $-1/\tilde{\chi}_2^{\,*}(\Omega_1+\wmod)$, with $\tilde{\chi}_n$ the optomechanically modified susceptibility that appears in \maineqref{eq:multimode:b1spectrum}. It then gives

\begin{equation}
  V_{12} \approx \tilde{\chi}_2^{\,*}(\Omega_1+\wmod)\!\left(-\sigma_{12}\,V_{22}
  - \sigma_{21}^*\,V_{11} + Q_{12}\right).
  \label{eq:SI:V12approx}
\end{equation}

Substituting $V_{12}^*$ from \cref{eq:SI:V12approx} into \cref{eq:SI:lyap11} and separating real and imaginary contributions:

\begin{equation}
  \begin{split}
  &-\Gtot{1} V_{11}
  + 2\,\mathrm{Re}\Bigl[
    \tilde{\chi}_2\bigl(
      |\sigma_{12}|^2\,V_{22}
      + \sigma_{12}\sigma_{21}\,V_{11}\\
      &\qquad - \sigma_{12}\,A_{\times,12}^*
    \bigr)
  \Bigr] + Q_{11} = 0,
  \end{split}
\end{equation}
where $\tilde{\chi}_2 = \tilde{\chi}_2(\Omega_1+\wmod)$.

The coefficient of $V_{11}$ defines the multimode cooling rate $\Gmulti = -2\,\mathrm{Re}[\sigma_{12}\sigma_{21}\,\tilde{\chi}_2(\Omega_1+\wmod)]$, which is \maineqref{eq:multimode:Gmulti}. It is positive in the enhanced-cooling regime, where the argument of $\sigma_{12}\sigma_{21}\tilde{\chi}_2$ has rotated past $-\pi/2$. Using $Q_{11} = \gamma_1 \nth{1} + A_{+,1}$ and $Q_{12} = A_{\times,12}$ from \cref{eq:SI:Qfull}, the solution for $V_{11} = \neff{1}$ is:

\begin{equation}
  \begin{split}
  \neff{1} = \frac{1}{\Gtot{1} + \Gmulti}\Bigl[&\left(\gamma_1 \nth{1} + A_{+,1}\right)
    + 2\,\mathrm{Re}(\tilde{\chi}_2)\,|\sigma_{12}|^2\,\neff{2}\\
  &- 2\,\mathrm{Re}\!\left(\tilde{\chi}_2\,\sigma_{12}\,A_{\times,12}^*\right)\Bigr].
  \end{split}
  \label{eq:SI:n1eff_full}
\end{equation}

The three terms are the single-mode heating of mode 1, the occupancy of mode 2 transmitted through the coupling, and the cross-backaction correction, and they are divided by the total damping of mode 1, $\Gtot{1}+\Gmulti$.

To see where this damping is largest, we write $\sigma_{12}\sigma_{21} = |\sigma_{12}\sigma_{21}|\,e^{i\varphi}$ and $\delta = \Omega_1 + \wmod - \Omega_2$ for the detuning of the modulation from the resonance, including the optical spring of mode 2 in $\Omega_2$, so that $\tilde{\chi}_2 = (\Gtot{2}/2 - i\delta)^{-1}$ and
\begin{equation}
  \Gmulti(\delta) = -2\,|\sigma_{12}\sigma_{21}|\,
  \frac{(\Gtot{2}/2)\cos\varphi - \delta\sin\varphi}{(\Gtot{2}/2)^2 + \delta^2}.
  \label{eq:SI:Gmulti_lineshape}
\end{equation}
It changes sign at $\delta = (\Gtot{2}/2)\cot\varphi$, and its cooling lobe peaks at $\delta^* = (\Gtot{2}/2)(1+\cos\varphi)/\sin\varphi$ with
\begin{equation}
  \Gmulti^{\max} = \frac{2\,|\sigma_{12}\sigma_{21}|}{\Gtot{2}}\,(1 - \cos\varphi).
  \label{eq:SI:Gmulti_max}
\end{equation}
The phase $\varphi$ follows from \maineqref{eq:multimode:sigmas}: on the resonance the cavity factors of $\sigma_{12}$ and $\sigma_{21}$ coincide pair by pair, so that a single pair gives $\varphi = 2\arg c_k$, independent of the tone phases, with $c_k = \chi(\Delta_k + \Omega_2) - \chi^*(\Delta_k - \Omega_1)$ the cavity response at the two photons the pair shares. At $\Delta_0 = -\Omega_2$ the shared anti-Stokes photon of the $(0,+1)$ pair is cavity-resonant and its shared Stokes photon is the one that sets $\epsilon$, so that $\varphi \approx -2\arctan[\kappa/2(\Omega_1+\Omega_2)]$ and $2/(1-\cos\varphi) \approx 1/\epsilon$. For $\varphi = \pi$ the lobe is a damping peak on the resonance, for $\varphi = \pm\pi/2$ the cooling and anti-damping lobes are comparable and $\Gmulti$ vanishes on the resonance, as in \mainfigref[b and c]{fig:res-theory:scheme-freq-space}, and for $\varphi \to 0$ the resonance is an anti-damping dip with only a weak cooling lobe in its wing. At $14.0\uW$, $\beta = 0.45\pi$ and $\Delta_0 = -\Omega_2$ we find $\varphi \approx -0.15\pi$ on this branch. At the lobe maximum the prefactor of $\neff{2}$ in \cref{eq:SI:n1eff_full}, $2\,\mathrm{Re}(\tilde{\chi}_2)\,|\sigma_{12}|^2/\Gmulti$, equals $|\sigma_{12}|/|\sigma_{21}|$ for any $\varphi$, and we evaluate the floor there.

To find the floor that the scheme approaches, we now take the limit of large multimode coupling, $\Gmulti \gg \Gtot{1}$, in which \cref{eq:SI:n1eff_full} reduces to the transferred occupancy of mode 2 plus the part of the heating of mode 1 that grows with the drive. The thermal term $\gamma_1\nth{1}$ is independent of the drive and becomes negligible relative to the denominator. The optical term $A_{+,1}$ does not: it scales with the same tone powers as $\Gmulti$. To see this, and to find the ratio $A_{+,1}/\Gmulti$ that therefore remains in the floor, consider a single adjacent tone pair, with the mode-1 tone of coupling $G_{11} = g_1|\alpha_{k+1}|$ near the cavity resonance, $|\Delta_{k+1}| = \Omega_1 \ll \kappa$, and the auxiliary tone of coupling $G_{22} = g_2|\alpha_k|$ on its anti-Stokes resonance $\Delta_k = -\Omega_2$. Then $|\sigma_{12}| \approx |\sigma_{21}| \approx 2G_{11}G_{22}/\kappa$, and at the lobe maximum, \cref{eq:SI:Gmulti_max},

\begin{equation}
  \Gmulti^{\max} \approx \frac{8\,G_{11}^2 G_{22}^2}{\kappa^2\,\Gtot{2}}\,(1 - \cos\varphi)
  = \frac{4G_{11}^2}{\kappa}\,\frac{\Gopt{2}}{\Gtot{2}}\,\frac{1 - \cos\varphi}{2},
  \label{eq:SI:Gmulti_saturation}
\end{equation}
using $\Gopt{2} = 4G_{22}^2/\kappa$. The auxiliary power cancels: mode 1 cannot be damped faster than the rate at which its own tone scatters photons, and the auxiliary mode and the phase $\varphi$ only determine which fraction of that rate ends up as dissipation. This is the effective linewidth $\tilde\gamma_1 = 4|G|^2/\kappa$ of Ref.~\cite{ojanenGroundstateCoolingMechanical2014}. Since mode 1 is sideband-unresolved, its Stokes rate from the same tone is $\approx A_{-,1} \approx 4G_{11}^2/\kappa$. The auxiliary tone scatters off mode 1 as well: its Stokes sideband on mode 1 lies at $\Delta_k - \Omega_1 \approx -\Omega_2$, where the cavity response is suppressed only by the factor $\epsilon$ of \cref{eq:SI:epscross}, so that $A_{+,1} \approx 4G_{11}^2/\kappa + \epsilon\,4G_{12}^2/\kappa$ with $G_{12} = g_1|\alpha_k|$ the coupling of the auxiliary tone to mode 1, and

\begin{equation}
  \begin{split}
  \frac{A_{+,1}}{\Gmulti^{\max}} &\approx \frac{\Gtot{2}}{\Gopt{2}}\,\frac{2}{1 - \cos\varphi}\left(1 + \epsilon\,\frac{|\alpha_k|^2}{|\alpha_{k+1}|^2}\right)\\
  &= \left(1 + \frac{\gamma_2}{\Gopt{2}}\right)\frac{2}{1 - \cos\varphi}\left(1 + \epsilon\,\frac{|\alpha_k|^2}{|\alpha_{k+1}|^2}\right)\\
  &\geq 1.
  \end{split}
  \label{eq:SI:Aplus_lock}
\end{equation}
The leading factors are independent of both drive powers, and $2/(1-\cos\varphi)$ is one only for $\varphi = \pi$. The correction is not: $\Gmulti$ saturates with the auxiliary power by \cref{eq:SI:Gmulti_saturation}, whereas the scattering of the auxiliary tone off mode 1 keeps growing, so the auxiliary power has an optimum rather than dropping out. At the reference point of the main text ($14.0\uW$, $\beta = 0.45\pi$, $\Delta_0 = -\Omega_2$) the auxiliary tone is the weaker of the two in the cavity, $|\alpha_0|/|\alpha_{+1}| \approx 0.3$, since the mode-1 tone sits close to the cavity resonance. The correction is below one percent; it matters only for the asymptotic statements of \cref{subsec:SI:coolinglimit-sum}. The mode-2 term $2\,\mathrm{Re}(\tilde{\chi}_2)|\sigma_{12}|^2 \neff{2}$ scales like $\Gmulti$, so their ratio is fixed at a given operating point. The cross-diffusion term is suppressed on this branch for red-detuned tones by the factor $\epsilon$ of \cref{eq:SI:epscross}, and we neglect it in this limiting analysis. Evaluating at the maximum of the cooling lobe:

\begin{equation}
  \begin{split}
  \neff{1} &\xrightarrow{\;\Gmulti \gg \Gtot{1}\;}
  \frac{A_{+,1}}{\Gmulti^{\max}} + \frac{|\sigma_{12}|}{|\sigma_{21}|}\,\neff{2}\\
  &\;\geq\; \frac{\Gtot{2}}{\Gopt{2}}\,\frac{2}{1-\cos\varphi} + \frac{|\sigma_{12}|}{|\sigma_{21}|}\,\neff{2},
  \end{split}
  \label{eq:SI:coolingfloor}
\end{equation}
where the inequality follows from \cref{eq:SI:Aplus_lock}. The second term transfers the auxiliary occupancy to the target mode. Even once mode 2 is itself backaction-limited, this does not reach $\nqba{2} = (\kappa/4\Omega_2)^2$, because the mode-1 tone keeps heating mode 2, as discussed after \cref{eq:SI:V22approx}. The first term is the residual Stokes heating of the target mode, which the beamsplitter interaction cannot remove: it conserves the total phonon number, so a phonon created in mode 1 by a Stokes event can only be transferred to mode 2 and dissipated there, at best at the rate $\Gmulti$, and the balance leaves at least of order one phonon in mode 1. Since $A_{+,1}/\Gmulti$ is smallest at the lobe maximum and the transfer term is never negative, this bound holds at every modulation frequency, not only at the lobe maximum. At this carrier detuning the difference-frequency branch therefore lowers the floor from $\nqba{1} = \kappa/4\Omega_1$ at best to the order-one level, but not into the ground state, and it reaches that level only where $\varphi$ is close to $\pi$: at $\Delta_0 = -\Omega_2$, where $\varphi \approx -0.15\pi$, the factor $2/(1-\cos\varphi) \approx 18$ makes this branch cool poorly. Once the transfer term dominates, the optimum over $\wmod$ moves somewhat beyond the lobe maximum, where the prefactor of $\neff{2}$ falls below $|\sigma_{12}|/|\sigma_{21}|$ at the cost of a larger Stokes residual.

At the lobe maximum the prefactor is exactly one whenever the optomechanically mediated coupling is reciprocal, $|\sigma_{12}| = |\sigma_{21}|$, and the auxiliary occupancy is then transferred in full. With \cref{eq:SI:coolingfloor}, and $\Gtot{2}/\Gopt{2} = 1 + \gamma_2/\Gopt{2}$, the floor of the difference-frequency branch is then
\begin{equation}
  \neff{1} \;\geq\; \neff{2} + \left(1 + \frac{\gamma_2}{\Gopt{2}}\right)\frac{2}{1-\cos\varphi} \;\geq\; \neff{2} + 1,
  \label{eq:SI:difffloor}
\end{equation}
so the multimode coupling replaces the QBA limit $\nqba{1}$ of the target mode by the occupancy of the auxiliary mode, plus a Stokes residual of at least one phonon. Independent control of the tone amplitudes and phases, for instance through combined amplitude and phase modulation, allows $|\sigma_{12}| < |\sigma_{21}|$, a nonreciprocal interaction. Its effect on the floor is quantified in \cref{subsec:SI:nonreciprocity}.

\subsection{Sum-frequency branch}
\label{subsec:SI:coolinglimit-sum}

\Cref{eq:SI:coolingfloor} was derived for the difference-frequency resonance, where the interaction is of beamsplitter type. Most of that derivation carries over to the sum-frequency resonance of \cref{subsec:SI:sumfreq}, but we will now show that the floor it sets is different.

At their respective resonances the two dynamical matrices share the same structure. Setting $\wmod = \Omega_2 - \Omega_1$ in \maineqref{eq:multimode:Drot} makes the bare diagonal frequencies of both entries equal to $\bar\Omega = (\Omega_1+\Omega_2)/2$, and setting $\wmod = \Omega_1 + \Omega_2$ in \cref{eq:SI:Drotsumfreq} makes both equal to $\bar\Omega = (\Omega_1-\Omega_2)/2$. Absorbing $\mathrm{Re}\,\sigma_{nn}$ into $\Gtot{n}$ and neglecting the differential optical spring shift as in \cref{subsec:SI:coolinglimit}, either case reduces to

\begin{equation}
  D^\mathrm{rot} =
  \begin{pmatrix}
    -i\bar\Omega - \Gtot{1}/2 & -A \\
    -B & -i\bar\Omega - \Gtot{2}/2
  \end{pmatrix},
  \label{eq:SI:genericDrot}
\end{equation}
with $(A,B) = (\sigma_{12},\sigma_{21})$ or $(\sigma_{12}^s,\sigma_{21}^s)$. Note that $\sigma_{21}^s$ enters unconjugated: by $\chi(-\omega) = \chi^*(\omega)$, \cref{eq:SI:sumfreqsigmas} already defines it as the conjugate of the coefficient of $\hat{b}_1^\dagger$ in the equation of motion for $\hat{b}_2$.

Everything set by the eigenvalues of \cref{eq:SI:genericDrot} is therefore common to both resonances, including the hybridized linewidths and frequencies, the stability boundaries and the reversal mechanism of the multimode damping. Repeating the elimination of \cref{subsec:SI:coolinglimit} gives $\Gmulti = -2\,\mathrm{Re}(AB\,\tilde{\chi}_2)$ and a numerator contribution $2\,\mathrm{Re}(\tilde{\chi}_2)|A|^2\,V_{22}$ in both cases, which is why \mainfigref[b]{fig:res-exp:freq-lw-n1eff-mainsweep} shows sign switches of the same character at all four resonances.

There are however three differences. First, the detuning entering the eliminated equation is $\Omega_1 + \wmod - \Omega_2$ for the difference frequency but $\Omega_1 + \Omega_2 - \wmod$ for the sum frequency. The reversed sign of $\wmod$ makes the argument of the auxiliary-mode susceptibility sweep in the opposite sense, so enhanced cooling lies below the difference-frequency resonance and above the sum-frequency one.

Second, the dynamical variable is $\hat{b}_2^\dagger$, so that $V_{22}^s = \neff{2}+1$ and the diffusion element $Q_{22}^s$ of \cref{eq:SI:Q22sum} takes the place of $Q_{22}$. The coupling term in the $(2,2)$ equation likewise changes character: the two-mode-squeezing channel creates phonon pairs spontaneously, so it feeds mode 2 at a rate of order $\Gmulti(\neff{1}+1)$ rather than $\Gmulti\neff{1}$. This adds about $\Gmulti/\Gtot{2}$ to $\neff{2}$ even for an empty target mode.

Third, the cross-diffusion can no longer be neglected. On the difference-frequency branch it was suppressed by the factor $\epsilon$ of \cref{eq:SI:epscross}, but on the sum-frequency branch the shared vacuum photon is cavity-resonant, and by \cref{eq:SI:noisecommutator} $Q_{12}^s = A_{\times,12}^s \approx \sigma_{12}^s - \sigma_{21}^{s*}$ is as large as the coupling itself.

Propagating $V_{22}^s$ and $A_{\times,12}^s$ through \cref{subsec:SI:coolinglimit}, the cross-diffusion term of \cref{eq:SI:n1eff_full} becomes $-2\,\mathrm{Re}(\tilde{\chi}_2\sigma_{12}^s A_{\times,12}^{s*}) = -2\,\mathrm{Re}(\tilde{\chi}_2)|\sigma_{12}^s|^2 - \Gmulti$. Its first part cancels the $+1$ of $V_{22}^s$ in the transfer term, and its second part removes one phonon's worth of $\Gmulti$ from the numerator, at any modulation frequency. Evaluating at the maximum of the cooling lobe as in \cref{eq:SI:coolingfloor},

\begin{equation}
  \neff{1} \xrightarrow{\;\Gmulti \gg \Gtot{1}\;}
  \frac{A_{+,1}}{\Gmulti^{\max}} - 1 + \mathcal{T}\,\neff{2} + \frac{\gamma_1\nth{1}}{\Gmulti^{\max}},
  \quad
  \mathcal{T} = \frac{|\sigma_{12}^s|}{|\sigma_{21}^s|},
  \label{eq:SI:coolingfloorsum}
\end{equation}
with the same transfer prefactor $\mathcal{T}$ as before and the thermal term of mode 1 retained, since it dominates at our bath temperature. Had the cross-diffusion been neglected, as on the difference-frequency branch, the result would have been $\mathcal{T}(\neff{2}+1)$ plus the Stokes residual, so that even a ground-state auxiliary mode would leave $\neff{1} \geq 1$. The vacuum noise instead interferes through the coupling and removes both the $+1$ of the auxiliary mode and the Stokes heating of the target mode by its own tone. The $-1$ is exact and independent of the reciprocity of the coupling, since \cref{eq:SI:noisecommutator} holds for any tone configuration; only the $\epsilon$-suppressed anti-normally ordered remainder has been dropped. On this branch the cavity factor of $\sigma_{21}^s$ is minus that of $\sigma_{12}^s$, so a single pair gives $\varphi = \pi + 2\arg[\chi(\Delta_k + \Omega_1) - \chi^*(\Delta_k + \Omega_2)]$, close to $\pi$ since the shared photon at $\Delta_0 + \Omega_2$ is cavity-resonant;

\begin{equation}
  \neff{1} \;\longrightarrow\; \neff{2} + \frac{\gamma_2}{\Gopt{2}}
  + \epsilon\,\frac{|\alpha_k|^2}{|\alpha_{k+1}|^2}\left(1 + \frac{\gamma_2}{\Gopt{2}}\right)
  + \frac{\gamma_1\nth{1}}{\Gmulti}.
  \label{eq:SI:sumfloorresult}
\end{equation}
What remains is the finite cooperativity of the auxiliary mode and the Stokes scattering of the auxiliary tone off the target mode. The physical picture is that the Stokes photon of the mode-1 tone and the anti-Stokes photon of the auxiliary tone on mode 2 are emitted into the same cavity frequency, so the amplitude for creating a phonon in mode 1 interferes destructively with the amplitude for annihilating one in mode 2, the two processes that the two-mode-squeezing interaction pairs. Equivalently, the auxiliary mode dresses the cavity with a transparency window at the Stokes frequency of mode 1, which is the optomechanically induced transparency picture of Ref.~\cite{ojanenGroundstateCoolingMechanical2014}. Their drive configuration --- one beam at $-\Omega_1$ and one at $-\Omega_2 - 2\Omega_1$ --- is a two-tone realization of this branch, and \cref{eq:SI:sumfloorresult} reproduces their result that the target mode inherits the occupancy of the auxiliary mode, with the effective linewidth of \cref{eq:SI:Gmulti_saturation}.

The sum-frequency branch is thus the one whose floor follows the auxiliary occupancy, while the difference-frequency branch retains, for the tone geometry used here, the order-one residual of \cref{eq:SI:coolingfloor}. At our bath temperature both floors are dominated by the thermal terms. At $14.0\uW$, $\beta = 0.45\pi$ and $\Delta_0 = -\Omega_2$, the full solution gives $\neff{1} = 34$ on the sum-frequency branch, close to the sum of the terms of \cref{eq:SI:coolingfloorsum}, $\mathcal{T}\neff{2} = 0.92 \times 18$, $\gamma_2/\Gopt{2} = 1.6$ and $\gamma_1\nth{1}/\Gmulti = 17$. On the difference-frequency branch at the same detuning  (which cools the auxiliary mode optimally but places the enhanced-cooling lobe in the wing of its resonance), $\neff{1}$ stays above $250$. The branches separate once $\Gopt{2} \gg \gamma_2\nth{2}$ and $\Gmulti \gg \gamma_1\nth{1}$, with a stronger carrier, a colder bath or an auxiliary mode of higher mechanical $Q$.

\subsection{Nonreciprocal coupling}
\label{subsec:SI:nonreciprocity}

On both branches the auxiliary occupancy reaches the target through the prefactor $\mathcal{T} = |\sigma_{12}|/|\sigma_{21}|$, which is one for reciprocal coupling. Here we ask when the Floquet drive makes the coupling nonreciprocal, $|\sigma_{12}| < |\sigma_{21}|$, and how far that lowers the floor.

A single adjacent tone pair gives a reciprocal coupling: at $\wmod = \Omega_1 + \Omega_2$ the resonant part of pair $k$ in \cref{eq:SI:sumfreqsigmas} is $\sigma_{12}^s = -g_1g_2\,\alpha_k^*\alpha_{k+1}\,\chi_k^*$ and $\sigma_{21}^s = g_1g_2\,\alpha_k\alpha_{k+1}^*\,\chi_k^*$, with $\chi_k = \chi(\Omega_2 + \Delta_k)$ the cavity response at the shared photon, so that $|\sigma_{12}^s| = |\sigma_{21}^s|$ whatever the phase of $\chi_k$; the same holds on the difference-frequency branch. Nonreciprocity requires interference between the two pairs, the mechanism by which nonreciprocal coupling between two membrane modes was realized in Refs.~\cite{xuNonreciprocalControlCooling2019, jiangNonreciprocalDynamicsCryogenic2020}. With $P_k = \alpha_k^*\alpha_{k+1}$,

\begin{equation}
  |\sigma_{12}^s|^2 - |\sigma_{21}^s|^2
  = -4\,(g_1g_2)^2\;\mathrm{Im}\!\left(P_0 P_{-1}^*\right)\,\mathrm{Im}\!\left(\chi_0^*\chi_{-1}\right),
  \label{eq:SI:nonrecip_condition}
\end{equation}
which needs a phase difference between the two shared photons to be significant. This is supplied by the cavity filter since the two tones are $\Omega_1 + \Omega_2$ apart. Moreover, tone phases should not be equally spaced: $\arg\alpha_{-1} + \arg\alpha_{+1} \neq 2\arg\alpha_0 \pmod{\pi}$. Pure phase modulation fixes the input phases, $J_{-1} = -J_1$, but the cavity adds $\arg\chi(\Delta_{-1}) + \arg\chi(\Delta_{+1}) - 2\arg\chi(\Delta_0) \approx 0.4\pi$ at $\Delta_0 = -\Omega_2$, so the three-tone drive is already mildly nonreciprocal in the favorable direction: $|\sigma_{12}^s|/|\sigma_{21}^s| \approx 0.9$ at the reference point of the main text. The effect stays small because the filter also fixes $|\alpha_{-1}|/|\alpha_{+1}| \approx 0.13$. Combined amplitude and phase modulation, which sets the two first-order sidebands independently, or a separate laser for the $k = -1$ tone, would supply the missing control over $|\alpha_{-1}|$ and its phase in order to make nonreciprocity large.

The gain available from this control follows from \cref{eq:SI:coolingfloorsum}. The $(-1,0)$ pair, whose shared photon at $-\Omega_1 - \Omega_2$ is far from resonance, adds to the resonant, dissipative coupling of the $(0,+1)$ pair a predominantly coherent two-mode-squeezing coupling, so that $\sigma_{12}^s \propto 1 + it e^{i\psi}$ and $\sigma_{21}^s \propto 1 + it e^{-i\psi}$, with relative strength $t \approx 0.4\,|\alpha_{-1}|/|\alpha_{+1}|$ and phase $\psi = 2\arg\alpha_0 - \arg\alpha_{-1} - \arg\alpha_{+1}$; pure phase modulation gives $t \approx 0.05$. A coherent two-mode-squeezing coupling causes anti-damping, so it reduces $\Gmulti$ by the factor $1 - t^2$ and raises the Stokes residual and the thermal term of \cref{eq:SI:coolingfloorsum} by $1/(1-t^2)$, but at $\psi = \pi/2$ it reduces the transfer coefficient faster, $\mathcal{T} = |\sigma_{12}^s|/|\sigma_{21}^s| = (1-t)/(1+t)$. On resonance, for negligible thermal load on mode 1 and to leading order in $t$,

\begin{equation}
  \neff{1} \;\longrightarrow\;
  \frac{1 + \gamma_2/\Gopt{2}}{1 - t^2} - 1
  + \frac{1-t}{1+t}\,\neff{2},
  \label{eq:SI:nonrecip_floor}
\end{equation}
which is minimized at $t\,(1 + \gamma_2/\Gopt{2}) = \neff{2}(1-t)^2$, giving approximately $\sqrt{(1 + \gamma_2/\Gopt{2})\,\neff{2}} - 1$ for $\neff{2} \gg 1$ in place of $\neff{2} + \gamma_2/\Gopt{2}$, and a gain of order $(\neff{2})^2$ for $\neff{2} \ll 1$. The expansion holds for $t \lesssim 0.2$, and the optimum lies beyond it: for our device parameters, with $|\alpha_{-1}|$ and its phase free and the thermal load on mode 1 suppressed, the full solution of the Lyapunov equation lowers the floor by about a third, from $11.6$ to $7.9$ phonons, because $\mathcal{T}$ saturates near $0.55$. Nonreciprocity therefore lowers the floor while the auxiliary mode is thermally limited, as in the present experiment, but not once it is in its ground state, where $\gamma_2/\Gopt{2}$ and the $\epsilon$ terms remain. The favorable phase is close to that of pure phase modulation and the unfavorable one to that of amplitude modulation, so what the present scheme lacks in order to induce significant nonreciprocity is control over the amplitude of the $k = -1$ sideband rather than its phase.

\section{Photothermal forces}
\label{sec:SI:photothermal}
The optomechanical platform used in this work is known to support photothermal backaction alongside radiation-pressure backaction\cite{primoAccurateModelingCharacterization2021,jansenControllingPhotothermalForces2026}. Here we include it in the multimode model of \cref{sec:SI:multimode}, following the model of Ref.~\cite{primoAccurateModelingCharacterization2021}. It is not implemented in the numerical sweeps of the main text, because accurate values of the photothermal coupling rate are hard to obtain.

The temperature field of the device is decomposed into thermal eigenmodes $\tilde{T}_k(\mathbf{r})$ with decay times $\tau_k$, so that $\delta T(\mathbf{r},t) = \sum_k \delta\theta_k(t)\tilde{T}_k(\mathbf{r})$, and the amplitude of the $k$-th thermal eigenmode is driven by the fluctuations of the absorbed power\cite{primoAccurateModelingCharacterization2021},

\begin{equation}
  \dot{\delta\theta}_k
  = -\frac{1}{\tau_k}\delta\theta_k
  + \frac{\hbar\omega_0\kappa_\mathrm{abs} R_k^\theta}{\tau_k}
  \bigl(\alpha^*(t)\hat{d}(t) + \alpha(t)\hat{d}^\dagger(t)\bigr),
  \label{eq:SI:thermal_eom}
\end{equation}
where $\kappa_\mathrm{abs}$ is the optical absorption rate and $R_k^\theta$ the thermal resistance of thermal mode $k$, an overlap integral of the optical energy density with the thermal eigenmode profile that carries no mechanical mode index. The thermal modes act back on mechanical mode $n$ through thermoelastic stress, with coupling $\Lambda_{k,n}^\theta = \int S_n^x\mathbin{:} (\mathbf{c}\mathbin{:}\boldsymbol{\alpha}_\mathrm{th})\tilde{T}_k\,dV$, where $S_n^x(\mathbf{r})$ is the normalized strain profile of mode $n$, $\mathbf{c}$ the stiffness tensor and $\boldsymbol{\alpha}_\mathrm{th}$ the thermal expansion tensor. This adds the force $(i x_{\mathrm{zpf},n}/\hbar)\sum_k \Lambda_{k,n}^\theta\,\delta\theta_k$, with $x_{\mathrm{zpf},n} = \sqrt{\hbar/2m_{\mathrm{eff},n}\Omega_n}$, to the equation of motion of $\hat{b}_n$ in \maineqref{eq:multimode:EOM}. We neglect the thermo-optic shift of the cavity resonance by the temperature oscillation at $\Omega_n$; the static heating shifts it too, but enters only through the detuning, which we take from the heated resonance (\mainsecref{subsec:meth:detuning}). The photothermal loop is therefore one-directional: the photothermal force adds to the optical force on each mode but not to the transduction of its motion into the light, and the substitution $g_n \to g_n + g_{\mathrm{pt},n}$ everywhere would be incorrect.

Since $\kappa \gg \Omega_1, 1/\tau_k$, the optical field is adiabatically eliminated as before, and eliminating $\delta\theta_k[\omega]$ with the thermal susceptibility $\chi_k^\theta(\omega) = (1/\tau_k - i\omega)^{-1}$ turns the thermoelastic force into a force proportional to the photon-number fluctuation $\delta\hat{n}$, with the effective photothermal coupling\cite{primoAccurateModelingCharacterization2021}

\begin{equation}
  G_{\mathrm{pt},n}(\omega)
  = \omega_0\kappa_\mathrm{abs}
  \sum_k \frac{R_k^\theta\,\Lambda_{k,n}^\theta\,\chi_k^\theta(\omega)}{\tau_k}.
  \label{eq:Gpt}
\end{equation}
With $g_{\mathrm{pt},n} = -G_{\mathrm{pt},n}(\Omega_n)\, x_{\mathrm{zpf},n}$, where the minus sign accounts for the radiation-pressure force per photon being $-\hbar g_n/x_{\mathrm{zpf},n}$ in our convention while the photothermal force per photon is $+\hbar G_{\mathrm{pt},n}$\cite{primoAccurateModelingCharacterization2021}, mode $n$ is driven by $-i(g_n + g_{\mathrm{pt},n})\,\delta\hat{n}$. Because of the one-directional loop, every optical term in the equation of motion of mode $n$, and only there, is thus multiplied by

\begin{equation}
  u_n = 1 + \frac{g_{\mathrm{pt},n}}{g_n}.
  \label{eq:SI:ptfactor}
\end{equation}
For the diagonal self-energy, $\sigma_{nn} \to u_n\sigma_{nn}$ is the three-tone extension of the single-tone photothermal self-energy of Ref.~\cite{primoAccurateModelingCharacterization2021}, with $|a_0|^2$ replaced by $\sum_k|\alpha_k|^2$. The coupling $\sigma_{12}$, which describes the force on mode 1 from the photon-number fluctuations that mode 2 imprints on the modulated drive, becomes $u_1\sigma_{12}$, since the thermal response belongs to the mode being driven. For the gigahertz mode all relevant thermal modes satisfy $\Omega_2\tau_k \gg 1$, so that $\chi_k^\theta(\Omega_2) \approx i/(\Omega_2\tau_k) \to 0$ and $u_2 = 1$: the thermal field cannot follow the gigahertz oscillation, and $\sigma_{21}$ and $\sigma_{22}$ are unchanged. In the dynamical matrix $D^\mathrm{rot}$, \maineqref{eq:multimode:Drot}, the photothermal force therefore replaces $\sigma_{11}$ by $u_1\sigma_{11}$ and $\sigma_{12}$ by $u_1\sigma_{12}$, and leaves the entries of mode 2 unchanged.

In the diffusion matrix $Q$, \maineqref{eq:multimode:Q}, the photothermal force filters the optical force noise on mode 1 by the same factor, $A_{+,1} \to |u_1|^2 A_{+,1}$, and $A_{\times,12}$ acquires one factor of $u_1$, so that it is still built from the adjacent-tone products of \cref{eq:SI:Across}. Absorption adds no noise of its own, and the heating it causes enters through a raised $T$ in the $\gamma_n\nth{n}$ terms. Because the self-energy carries one factor of $u_1$ and the noise two, the photothermal damping of mode 1 no longer comes with the Stokes heating that the fluctuation-dissipation relation of the cavity attaches to radiation-pressure damping, which is how it can pre-cool the mode.

The structure of \cref{eq:SI:coolingfloor} therefore survives, so that the asymptotic occupancy is still proportional to $\neff{2}$, but because only $\sigma_{12}$ acquires the factor $u_1$, the coupling becomes nonreciprocal by the factor $|u_1|$. From the ratio of the measured optical spring to the measured damping in the $9.4\uW$ single-tone sweep we estimate $|u_1|^2 \approx 1.4$, between $1.1$ and $2.0$ within the uncertainty of the photon number. This outweighs the favorable $|\sigma_{12}^s|/|\sigma_{21}^s| \approx 0.9$ of the radiation-pressure drive (\cref{subsec:SI:nonreciprocity}), so the net nonreciprocity is most likely in the unfavorable direction, which by \cref{eq:SI:coolingfloor} raises the floor, though only slightly. Photothermal backaction thus helps by pre-cooling mode 1, through the additional damping $2\,\mathrm{Re}\,\sigma_{11}^\mathrm{pt}$, but does not improve the asymptotic multimode limit.

Our data do not resolve this change. At the operating point of \mainfigref[c]{fig:res-exp:lw-n1eff-approach_spectra}, the thermal inflow $\gamma_1\nth{1}$ accounts for $29$--$39$ of the $47 \pm 13$ phonons. The remainder is consistent with $\mathcal{T}\neff{2} \approx 17$--$20$, but too uncertain to separate the two cases. It does show that the excess inflow of the single-tone sweep (\cref{subsec:SI:flux}) is suppressed more than tenfold, as expected for a carrier close to the cavity resonance.

\section{Heating and calibration}
\label{sec:SI:heatingcalib}

This section collects the measurements behind the occupancies of the main text: the optical heating of the device and the temperature of the mechanical bath that it sets, the excess phonon inflow that limits single-tone cooling, the transduction of the probe, and the drive detuning and intracavity photon number with which we compare single-tone and multitone cooling at equal photon number. The occupancy and detuning calibrations themselves are summarized in the \mainsecref{subsec:meth:calib-occup}; here we give the supporting data and details.

\subsection{Optical heating}
\label{subsec:SI:expm-heating}

For an optical resonator with a resonance frequency $\omega_0$, the total fractional shift of the resonance frequency when the device is heated from a base temperature $T_0$ to a final temperature $T_f$ is obtained by the following integral\cite{kommaThermoopticCoefficientSilicon2012}:

\begin{equation}
  \begin{split}
  \frac{\Delta\omega}{\omega_0} &\equiv \frac{\omega(T_f) - \omega(T_0)}{\omega(T_0)}\\
  &= -\int_{T_0}^{T_f} \left[ \frac{1}{n_\mathrm{eff}}\frac{dn_\mathrm{eff}}{dT}(T') + \alpha(T') \right] dT',
  \end{split}
  \label{eq:SI:integral_shift}
\end{equation}

where $n_\mathrm{eff}$ is the effective modal refractive index, and $\alpha(T)$ is the linear thermal expansion coefficient. We approximate $n_\mathrm{eff}^{-1}\,dn_\mathrm{eff}/dT$ by the bulk value $n_\mathrm{Si}^{-1}\,dn_\mathrm{Si}/dT$ of silicon, with $n_\mathrm{Si} = 3.48$, which assumes that the optical mode is confined to the silicon. At cryogenic temperatures both $dn_\mathrm{eff}/dT$ and $\alpha$ vary by several orders of magnitude and cannot be treated as constants, hence the need for the integral.

The fractional shift $\Delta\omega/\omega_0$ is measured by comparing the probe resonance frequency acquired during the drive detuning sweep, $\omega_\mathrm{pumped}$, to a reference frequency $\omega_\mathrm{ref}$ obtained with the drive laser tuned far from the resonance, where the intracavity drive photon number is negligible and no significant heating occurs. The reference scans are taken immediately before and after the detuning sweep, together with the occupancy calibration (\mainsecref{subsec:meth:calib-occup}), and their mean is used as $\omega_\mathrm{ref}$. The fractional shift entering \cref{eq:SI:integral_shift} is then

\begin{equation}
  \frac{\Delta\omega}{\omega_0} \;=\; \frac{\omega_\mathrm{pumped} - \omega_\mathrm{ref}}{\omega_\mathrm{ref}}.
\end{equation}

\begin{figure*}[tbp]
  \centering
  \includegraphics{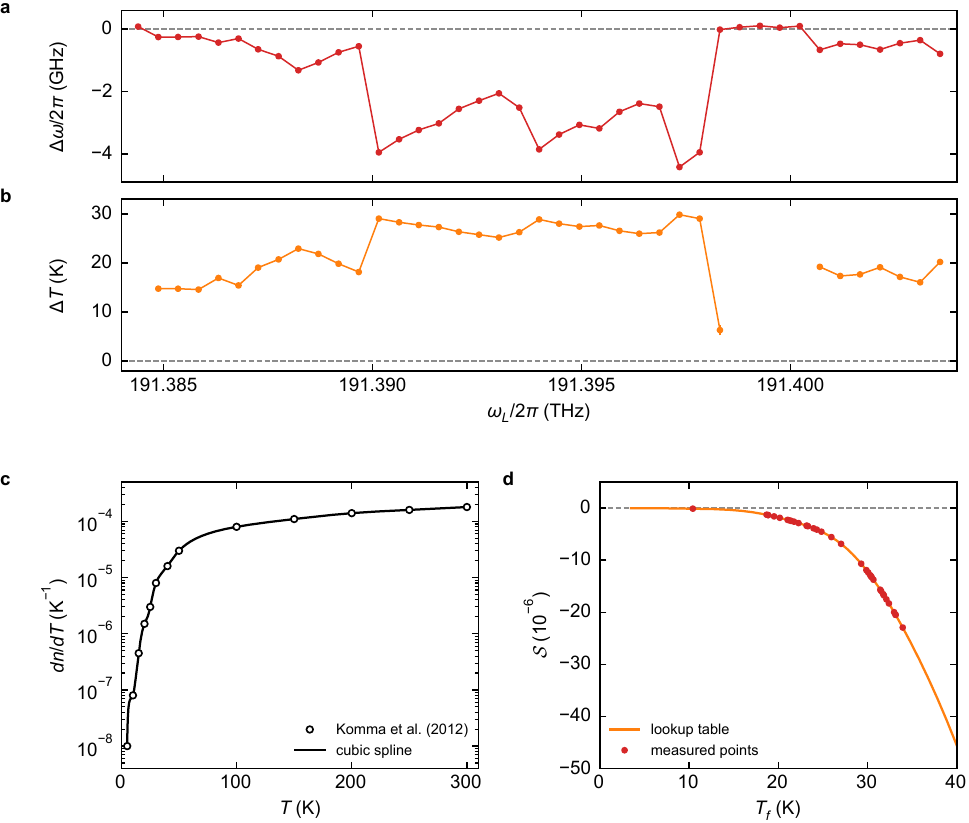}
  \caption{
    Estimate of the optical heating of the device.
    \subfigid{a} Measured probe mode shift $\Delta\omega/2\pi$ as a function of drive laser frequency $\omega_L/2\pi$, relative to the far-detuned reference.
    \subfigid{b} Corresponding device temperature rise $\Delta T$ obtained by inverting \cref{eq:SI:integral_shift}; scans with a blue shift, $\Delta\omega > 0$, cannot be inverted and are left out.
    \subfigid{c} Thermo-optic coefficient $dn/dT$ of silicon at $\lambda \approx 1550$ nm, digitized from Ref.~\cite{kommaThermoopticCoefficientSilicon2012} (circles), with the cubic spline used in the integral (line).
    \subfigid{d} Cumulative fractional shift $\mathcal{S}(T_f)$ from $T_0 = 4.1$ K, used as the lookup table for the inversion, with the measured shifts of \subfigid{a} at their inverted temperatures (dots).
    \label{fig:SI:optical-heating}
  }
\end{figure*}

For the thermo-optic coefficient of silicon at $\lambda \approx 1550\,\mathrm{nm}$ we use values digitized from Figure 2 of Ref. \cite{kommaThermoopticCoefficientSilicon2012}, which covers the range 5\,K-300\,K, shown in \cref{fig:SI:optical-heating}c. The coefficient rises from ${\approx}10^{-8}\,\mathrm{K}^{-1}$ at 5\,K to $1.8\times10^{-4}\,\mathrm{K}^{-1}$ at room temperature. The thermal expansion coefficient $\alpha(T)$ is taken from Table I of Ref. \cite{lyonLinearThermalExpansion1977}. Silicon exhibits anomalous negative thermal expansion between approximately 18\,K and 124\,K, with a minimum of $\alpha \approx -4.7\times10^{-7}\,\mathrm{K}^{-1}$ near 80\,K. Both datasets are interpolated by cubic splines. To invert \cref{eq:SI:integral_shift}, we pre-compute a lookup table $\mathcal{S}(T_f)$ by numerically evaluating the integral on a uniform grid of $10^4$ temperature points from $T_0 = 4.1\,\mathrm{K}$ to 300\,K. The resulting curve is shown in \cref{fig:SI:optical-heating}d. Although $\alpha$ is negative and partially offsets the thermo-optic contribution over the anomalous-expansion range, the term $n_\mathrm{eff}^{-1}\,dn_\mathrm{eff}/dT$ dominates throughout (exceeding $|\alpha|$ by a factor of approximately 30 at 50\,K and 70 at 100\,K), so $\mathcal{S}(T_f)$ remains monotonically decreasing in $T_f$. This guarantees a unique inversion: the device temperature at each drive detuning is recovered by linear interpolation of the inverse function $T_f(\mathcal{S})$.

The probe mode shift $\Delta\omega/2\pi$, determined from the data of \cref{fig:SI:probe-monitor}, is depicted in \cref{fig:SI:optical-heating}a, with the corresponding device temperature shift $\Delta T$ in \cref{fig:SI:optical-heating}b. We observe significant heating of around 30 K when the drive laser or one of its modulation sidebands is close to the resonance, falling off fairly quickly when the detuning is increased.

\subsection{Temperature of the mechanical bath}
\label{subsubsec:SI:mech-bath}
\begin{figure*}[tbp]
  \centering
  \includegraphics{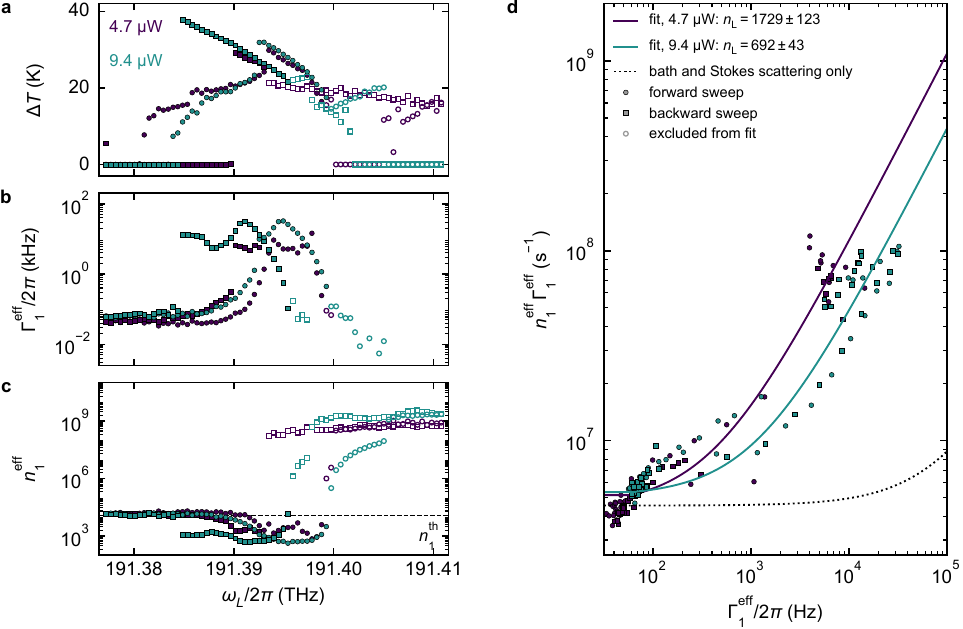}
  \caption{
    Single-tone sweeps at $4.7\uW$ (purple) and $9.4\uW$ (teal), forward (circles) and backward (squares).
    \subfigid{a} Device heating $\Delta T$ from the probe mode shift, as in \cref{fig:SI:optical-heating}b, \subfigid{b} linewidth $\Geff{1}/2\pi$, and \subfigid{c} occupancy $\neff{1}$ of mode 1 versus laser frequency $\omega_L/2\pi$. The dashed line in \subfigid{c} is $\nth{1}$ at $4.1$ K. Open markers are excluded from the fit in \subfigid{d} because the mode self-oscillates or the linewidth is outside the range of the fit ($5$ Hz to $350$ kHz).
    \subfigid{d} Phonon flux $\neff{1}\Geff{1}$ versus linewidth, with the fit of \cref{eq:SI:fluxfit} for each power (solid; $n_\mathrm{L} = 1729 \pm 123$ at $4.7\uW$ and $692 \pm 43$ at $9.4\uW$) and, dotted, the flux $\gamma_1\nth{1} + (A_{+,1}/\Gopt{1})\,\Gopt{1}$ with $A_{+,1}/\Gopt{1} = 6.7$ expected from the $4.1$ K bath and the Stokes scattering of the drive alone.
    \label{fig:SI:single-tone-series}
  }
\end{figure*}

The temperature of \cref{fig:SI:optical-heating}b is that of the optical mode volume, and the mechanical mode need not thermalize to it. We test this on two single-tone sweeps of the drive laser, at $4.7$ and $9.4\uW$, for which the device heating, mechanical linewidth, and occupancy are shown in \cref{fig:SI:single-tone-series}a-c. The heating follows the thermo-optic bistability of the drive mode: the forward sweep snaps onto the hot branch and the backward sweep releases from it several GHz lower, so the two directions trace a hysteresis loop. At the far-detuned end of each sweep the optical damping is negligible and the mode is a damped oscillator in equilibrium with its bath, so that its phonon flux $\neff{1}\Geff{1} = \gamma_1\nth{1}(T_\mathrm{bath})$ measures the bath temperature\cite{aspelmeyer2014review, meenehanSiliconOptomechanicalCrystal2014}. A linear fit of this bath temperature against the optical-mode heating $\Delta T$ at the same detunings gives $T_\mathrm{bath} = T_0 + 0.038\,\Delta T$. The mechanical bath thus rises by at most $1.4$\,K while the optical mode heats by up to $38$\,K. This is possible because the two modes sample different parts of the device: absorption heats the defect region in which the optical mode is localized (\mainfigref[e]{fig:intro}), while the flexural mode extends over the full length of the beams (\mainfigref[f]{fig:intro}), most of which stays close to the temperature of the clamped ends. This is unlike microresonators whose optical and mechanical modes share a volume and heat together\cite{macdonaldOptomechanicsThermometryCryogenic2016}, and similar to the absorption-induced phonon baths of silicon optomechanical crystals\cite{chanLaserCoolingNanomechanical2011, meenehanSiliconOptomechanicalCrystal2014, renTwodimensionalOptomechanicalCrystal2020, sonarHighefficiencyLownoiseOptomechanical2025, kolvikOptomechanicalCrystalLightresilient2026}. The thermal terms $\gamma_n\nth{n}$ of our model therefore see $4$--$6$\,K rather than the $\sim30$\,K that the optical mode reaches in the multitone measurement (\cref{fig:SI:optical-heating}b).

\subsection{Excess phonon flux of single-tone cooling}
\label{subsec:SI:flux}
The flux balance of \cref{subsubsec:SI:mech-bath} also quantifies the heating that accompanies the optical damping in the single-tone sweeps. In the bath model of Ref.\ \cite{meenehanSiliconOptomechanicalCrystal2014}, each bath coupled to the mode at rate $\gamma_i$ with occupancy $n_i$ contributes $\gamma_i$ to the damping and $\gamma_i n_i$ to the phonon inflow. For a single red-detuned tone the laser is such a bath, with damping $\Gopt{1}$ and occupancy $A_{+,1}/\Gopt{1}$ set by the Stokes rate of \cref{eq:SI:Aplus}\cite{marquardtQuantumTheoryCavityAssisted2007}. With the photothermal force of \cref{sec:SI:photothermal} included, which filters the force noise by $|u_1|^2$ and rotates the self-energy by $u_1$, this occupancy is $A_{+,1}/\Gopt{1} = 6.7$ for our device parameters. We fit
\begin{equation}
  \neff{1}\,\Geff{1} = \gamma_1\,\nth{1}(T_\mathrm{bath}) + n_\mathrm{L}\,\Gopt{1}
  \label{eq:SI:fluxfit}
\end{equation}
to all points of both sweep directions with a valid linewidth, with $T_\mathrm{bath}$ fixed point by point to the relation of \cref{subsubsec:SI:mech-bath} and the occupancy $n_\mathrm{L}$ of the optical channel as the only free parameter (\cref{fig:SI:single-tone-series}d). We find $n_\mathrm{L} = 1729 \pm 123$ at $4.7\uW$ and $692 \pm 43$ at $9.4\uW$, $100$--$250$ times the Stokes value; leaving $T_\mathrm{bath}$ free changes $n_\mathrm{L}$ by a few percent. A thermal bath cannot supply this inflow: at the single-tone optimum it would have to be hotter than the optical mode itself. An absorption-induced bath in the sense of Ref.\ \cite{meenehanSiliconOptomechanicalCrystal2014}, which adds its own damping $\gamma_p$ together with the inflow $\gamma_p n_p$, would also appear as a flux proportional to the excess linewidth, but is excluded by the multitone sweep, in which the flux at fixed carrier detuning does not follow the damping as $\wmod$ is varied. What the data establish is that the single-tone floor of $\neff{1} \approx 400$--$500$ in \cref{fig:SI:single-tone-series}c is predominantly set by this excess inflow, proportional to the optical damping and far above $\nqba{1} = 89$, and that it is present at both powers.

Its scaling with the optical damping, which for $\Omega_1 \ll \kappa$ is itself proportional to the slope of the intracavity photon number with detuning, together with its independence of the mechanical damping in the multitone sweep, is what a classical force from fluctuations of the cavity frequency relative to the laser produces: on the flank of the resonance such fluctuations are converted into photon-number fluctuations, which drive the mode through the same radiation-pressure and photothermal coupling without adding damping\cite{rablPhasenoiseInducedLimitations2009, safavi-naeiniLaserNoiseCavityoptomechanical2013}. Laser phase noise, thermorefractive noise of the cavity, and free-carrier dispersion\cite{krauseNonlinearRadiationPressure2015} all act in this way and differ only in their power dependence, which these two sweeps do not fix.

In summary, single-tone cooling in our device is limited neither by quantum backaction nor by the bath, but by this classical noise, which grows with the optical damping. The multimode damping $\Gmulti$ does not carry it, which is why the multitone scheme reaches an occupancy almost ten times lower at equal intracavity photon number (\cref{subsubsec:SI:ncav}).

\subsection{Transduction of the probe}
\label{subsec:SI:probe-monitor}
The occupancy calibration and its checks are described in the \mainsecref{subsec:meth:calib-occup}. Here we show how the transduction of the probe, which enters the calibration through the correction $(S_\mathrm{ref}/S)^2$, evolves during a sweep.

\begin{figure*}[tbp]
  \centering
  \includegraphics{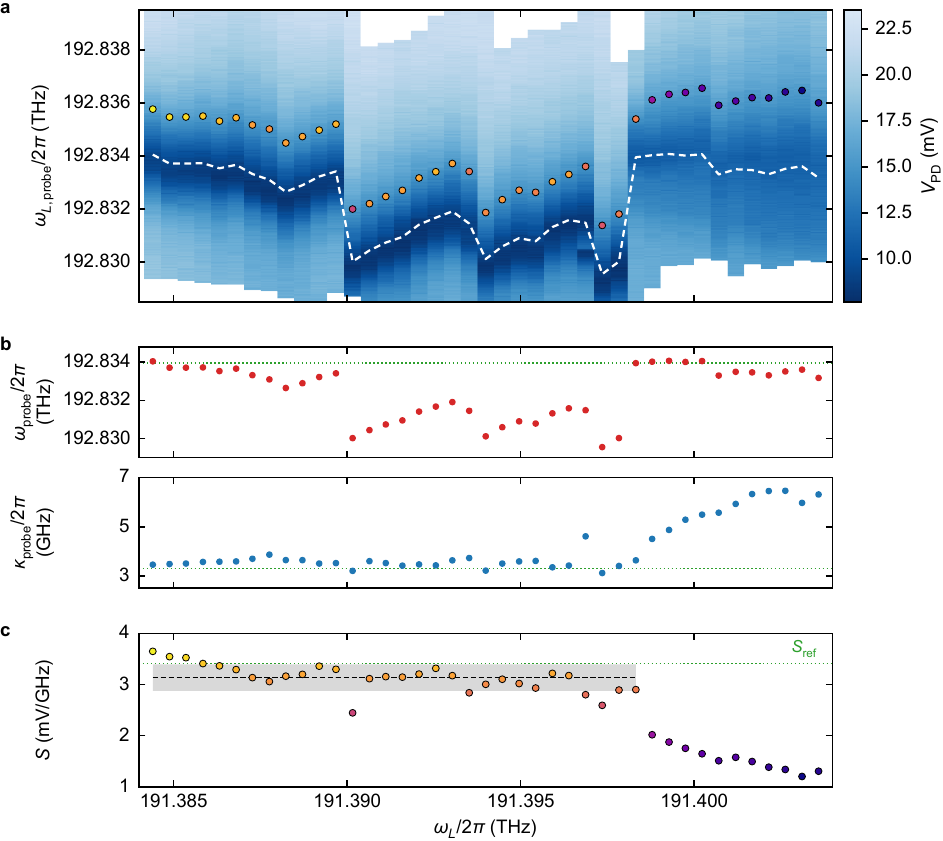}
  \caption{
    Monitoring the probe optical mode during the $14.0\uW$ sweep shown in \mainfigref{fig:res-exp:lw-n1eff-approach_spectra}.
    \subfigid{a} Probe reflection spectra, the photodiode voltage $V_\mathrm{PD}$ of the reflected probe against the probe laser frequency $\omega_{L,\mathrm{probe}}/2\pi$, as a function of drive laser frequency $\omega_L/2\pi$. Round markers indicate the PID setpoint, colored by the reflection derivative $S$ at the setpoint on the scale of \subfigid{c}. The dashed white line is the fitted probe resonance frequency.
    \subfigid{b} Fitted probe resonance frequency $\omega_\mathrm{probe}$ and linewidth $\kappa_\mathrm{probe}$ versus drive laser frequency. Dotted lines: their values in the two far-detuned reference scans.
    \subfigid{c} Reflection derivative $S$ at the setpoint, which sets the transduction. The dashed line and shaded band are the mean and standard deviation of $S$ before the onset of mechanical self-oscillation ($3.13 \pm 0.26$ mV/GHz, $8\%$); the dotted line is the far-detuned reference $S_\mathrm{ref}$ entering the correction $(S_\mathrm{ref}/S)^2$.
    \label{fig:SI:probe-monitor}
  }
\end{figure*}

The transduction of the probe does change during a sweep, and we monitor the shift and the transduction coefficients of the probe mode throughout. The transduction is proportional to the derivative of the reflected power of the probe in the limit of small mechanical displacement relative to the cavity linewidth. Therefore, we take a reflection spectrum of the probe mode and monitor how the reflection at the setpoint of the PID evolves during the measurement. The reflection spectra with setpoints indicated by round markers and slopes by the color of the marker, as taken during the $14.0\uW$ input power sweep from the main text, are shown in \cref{fig:SI:probe-monitor}a. By fitting the reflection spectra we can extract the optical resonance frequencies and linewidths of the probe mode. The resonance frequency is plotted as the dashed white line. Both are additionally shown in \cref{fig:SI:probe-monitor}b for easier identification of trends. Here we see that the linewidth changes significantly above a drive frequency of 191.3975 THz. This corresponds to the region where we start seeing mechanical self-oscillations. We therefore attribute this broadening to the mechanical modulation of the optical resonance. The effect of it on the transduction can be seen in \cref{fig:SI:probe-monitor}c, where we plot the slope at each drive laser frequency setting. Since the detected mechanical spectral area scales with the square of this slope, we correct every occupancy for the change in transduction relative to the far-detuned calibration, $\neff{1} \to \neff{1}\,(S_\mathrm{ref}/S)^2$, with $S$ the reflection derivative at the setpoint of the probe-dip scan taken at the same drive laser frequency and $S_\mathrm{ref}$ its mean value in the two far-detuned reference scans. For the $14.0\uW$ sweep the correction ranges from $0.87$ to $8.0$, and amounts to $1.21$ at the operating point shown in \mainfigref{fig:res-exp:lw-n1eff-approach_spectra}; for the $4.7\uW$ sweep it ranges from $0.73$ to $3.7$. The large values occur only in the blue-detuned self-oscillating regime, where the broadened probe resonance reduces the slope at the setpoint and the uncorrected areas would underestimate the mechanical energy several-fold. Away from self-oscillation the correction stays within about $20\%$, comparable to the propagated uncertainties of the main text.

\subsection{Drive detuning from the thermo-optic bistability}
\label{subsubsec:SI:pump-detuning}
The bistability model with which we calibrate the drive detuning and its fitted parameters are given in the \mainsecref{subsec:meth:detuning}. Here we describe the treatment of the drift of the probe resonance and the propagation of the calibration uncertainty. In the model the solution stays on its branch while that branch exists and jumps when it disappears, so the two switching frequencies alone already fix $\delta\omega_\mathrm{max}$ and a first estimate of the cold drive resonance $\omega_c^\mathrm{cold}$.

\begin{figure*}[tbp]
  \centering
  \includegraphics{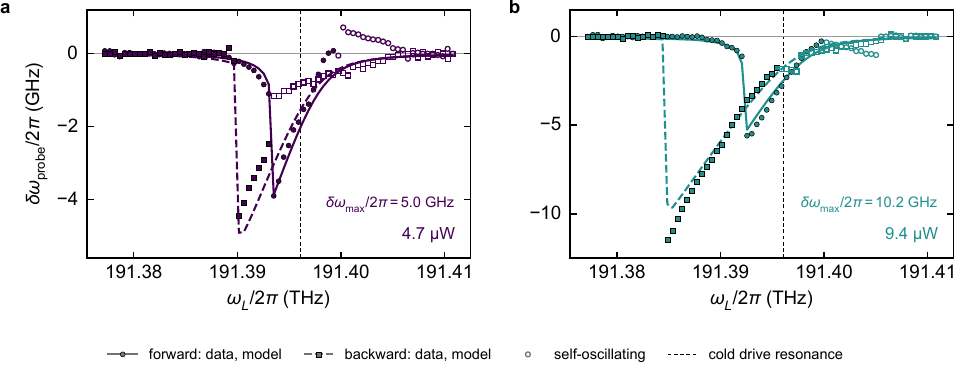}
  \caption{
    Calibration of the drive detuning from the thermo-optic bistability. Drift-corrected probe shift $\delta\omega_\mathrm{probe}/2\pi$ (negative: red shift) versus laser frequency $\omega_L/2\pi$ at \subfigid{a} $4.7\uW$ and \subfigid{b} $9.4\uW$, for the forward (circles) and backward (squares) sweeps. The lines are the bistability model of \maineqref{eq:meth:bistability}, with $\omega_c^\mathrm{cold} = \omega_c^{(0)} + \zeta\,\delta\omega_\mathrm{ref}$, fitted jointly to both panels with the parameters given in the text (solid: forward, dashed: backward). At the open markers the mechanical mode self-oscillates and broadens the probe mode; these points do not enter the fit. The dashed vertical line is the fitted cold drive resonance without probe drift, $\omega_c^{(0)}/2\pi = 191.39611$ THz.
    \label{fig:SI:pump-bistability}
  }
\end{figure*}

The probe shift is measured against the cold probe resonance, which slowly drifts by up to $1$\,GHz during a run and steps back blue during each heated excursion. We interpolate it through the unheated points of the sweeps and the reference traces. A common cold resonance for both sweep directions cannot reproduce both, as inside the hysteresis window the forward red shift exceeds the backward one by $1$\,GHz at the same laser frequency, which can be observed in \cref{fig:SI:pump-bistability}. The drive and probe modes are the even and odd supermodes of the nanobeam zipper cavity, so the drive cold resonance is allowed its own response to the drift $\delta\omega_\mathrm{ref}$ of the cold probe resonance, counted from its mean over the far-detuned scans before and after the $9.4\uW$ sweep, $\omega_c^\mathrm{cold} = \omega_c^{(0)} + \zeta\,\delta\omega_\mathrm{ref}$. A common-mode drift would give $\zeta = R$; the fit prefers $\zeta = -1.55$, i.e.\ the splitting of the supermodes changes while the device is hot, which takes care of the discrepancy between the sweep directions. This is unexpected at first sight, but could be due to the heating causing the stress in the device to change, in turn changing the gap between the beams.

With $R = \omega_\mathrm{drive}/\omega_\mathrm{probe} = 0.993$ fixed, the joint fit of $\omega_c^{(0)}$, $\zeta$, and one $\delta\omega_\mathrm{max}$ per power to both sweep directions at both powers is shown in \cref{fig:SI:pump-bistability}.

Refitting $\omega_c^{(0)}$ and $\zeta$ for every combination of $R \in \{0.89, 0.99, 1.09\}$, the three drift interpolations described next, and the sweep directions used brackets the systematic uncertainty: for the carrier of the operating point of \mainfigref{fig:res-exp:lw-n1eff-approach_spectra} it places $\Delta_0/2\pi$ between $-0.8$ and $2.1$\,GHz, about one cavity linewidth. A Monte Carlo simulation propagates the calibration uncertainty. Each of $4000$ samples draws $R \sim \mathcal{N}(0.993,\,0.05^2)$, truncated to $0.85 \le R \le 1.10$, and one of three equally likely ways to interpolate the probe drift across the heated excursions, where it cannot be measured: linearly, as a step at the start, or a step at the hottest point of the excursion. Each interpolation has its own best-fit $\zeta_\mathrm{fit}$ ($-1.55$, $-1.03$, $-1.50$), around which the sample draws $\zeta \sim \mathcal{N}(\zeta_\mathrm{fit},\,0.4^2)$, with $0.4$ about half the spread between fits to the two powers separately. With these values fixed, $\omega_c^{(0)}$ is refitted. Additionally propagating the noise of the probe shift gives the $1\sigma$ intervals $\zeta = -1.4 \pm 0.5$ and $\pm 0.4$\,GHz on $\omega_c^{(0)}$. For the carrier of the operating point this places $\Delta_0/2\pi$ between $1.3$ and $2.0$\,GHz, and the intracavity photon number of \maineqref{eq:meth:ncav} is uncertain by about $15\%$.

\subsection{Comparison at equal intracavity photon number}
\label{subsubsec:SI:ncav}
\begin{figure*}[tbp]
  \centering
  \includegraphics{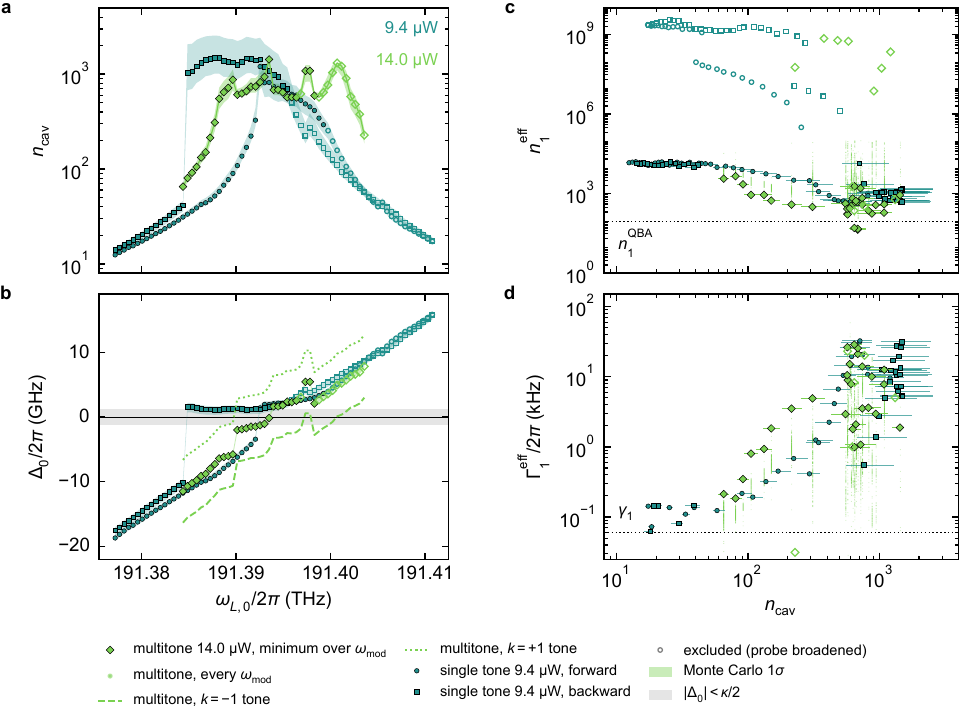}
  \caption{
    Occupancy and linewidth at equal intracavity photon number.
    \subfigid{a} Intracavity photon number from \maineqref{eq:meth:ncav} and \subfigid{b} carrier detuning $\Delta_0$ from the heated drive resonance, versus carrier frequency $\omega_{L,0}/2\pi$, for the $9.4\uW$ single-tone sweeps (teal; circles: forward, squares: backward) and the $14.0\uW$ multitone sweep (green diamonds), with the $k=-1$ and $k=+1$ tone detunings $\Delta_0 \mp \wmod$ as dashed and dotted lines. Colored bands are the $1\sigma$ Monte Carlo interval; the gray band in \subfigid{b} marks $|\Delta_0| < \kappa/2$.
    \subfigid{c} Occupancy $\neff{1}$ and \subfigid{d} linewidth $\Geff{1}/2\pi$ of mode 1 versus $n_\mathrm{cav}$, with the $1\sigma$ Monte Carlo interval of $n_\mathrm{cav}$ as horizontal bars. The diamond is the minimum over $\wmod$ for each multitone carrier and the small green points are all modulation frequencies; \subfigid{d} shows only linewidths within the fit range. Open markers are excluded because the mode self-oscillates such that the probe mode is broadened, significantly altering the transduction. The dotted lines are $\nqba{1} = \kappa/4\Omega_1$ in \subfigid{c} and $\gamma_1$ in \subfigid{d}.
    \label{fig:SI:ncav}
  }
\end{figure*}

With the detuning of every point known, the intracavity photon number follows from \maineqref{eq:meth:ncav} without free parameters, for the multitone sweep with the tones $k = -1, 0, +1$ of the modulation depth $\beta = 0.45\pi$ and for the single-tone sweeps with one tone. As a check, the heating maximum of the multitone sweep places the $k=-1$ tone $0.6$\,GHz blue of the hot resonance, which is where a forward sweep through the bistability heats most, just after the snap. \Cref{fig:SI:ncav}a,b show $n_\mathrm{cav}$ and the carrier detuning along the single-tone and multitone sweeps, with their $1\sigma$ Monte Carlo intervals as bands. The interval is asymmetric because $n_\mathrm{cav}$ depends nonlinearly on the detuning: near resonance any error in $\Delta$ lowers $n_\mathrm{cav}$, and on the flank an error towards resonance raises it more than an equal error away from it lowers it.

\Cref{fig:SI:ncav}c,d place both experiments on one axis. Sweeping $\wmod$ over its $30$\,MHz range changes $n_\mathrm{cav}$ of a carrier by only $1\%$, so all modulation frequencies of one carrier detuning lie in a vertical cloud, of which the diamond marks the lowest occupancy. The single-tone sweep reaches its lowest occupancy of $\neff{1} = 419 \pm 26$ at $n_\mathrm{cav} = 675$ ($497$--$847$), with a linewidth of $26$\,kHz. The multitone sweep reaches $\neff{1} = 47 \pm 13$ at $n_\mathrm{cav} = 685^{+112}_{-82}$, around a factor of ten lower at the same intracavity photon number, and the two neighbouring carrier detunings reach $44 \pm 16$ at $n_\mathrm{cav} = 684^{+111}_{-83}$ and $51 \pm 13$ at $n_\mathrm{cav} = 643^{+88}_{-61}$. At equal intracavity photon number, the multimode scheme therefore cools to an occupancy nearly a factor of ten lower.

\clearpage

\end{document}